\documentclass{article}
\usepackage{lineno}
\usepackage{siunitx}
\usepackage{pgfplots}
\pgfplotsset{compat=1.18}
\usepackage{ifthen}
\usepackage[margin=3cm]{geometry}
\usepackage{csvsimple-l3}
\usepackage{glossaries-extra}
\usepackage{mathtools}
\usepackage{amsmath}
\usepackage{amssymb}
\usepackage{biblatex}
\usepackage[export]{adjustbox}
\usepackage[ruled]{algorithm2e}
\usepackage{xurl}
\usepackage{csquotes}
\usepackage{microtype}
\usepackage[capitalize,noabbrev]{cleveref}
\usepackage{authblk}
\input{acronyms.gls}
\tikzset{font={\fontsize{6pt}{8}\selectfont}}
\usetikzlibrary{spy}
\newcommand{\drawcolorbar}{
	\pgfplotscolorbardrawstandalone[
		scale=0.32, colormap={blackwhite}{gray(0cm)=(0); gray(1cm)=(1)},
		colorbar horizontal, colorbar style={ticks=none},
	]%
}
\newcommand{\imwidth}{3cm}
\newcommand{\xpad}{.2cm}
\newcommand{\ypad}{2cm}

\newcommand{\Nleds}{N}
\newcommand{\ImagUnit}{\mathrm{j}}
\newcommand{\OneFunction}{1}

\newcommand{\FFDop}{\bb{D}}

\newcommand{\FT}[1]{\hat{#1}}
\newcommand{\Npatterns}{M}
\newcommand{\numapt}{\nu}
\newcommand{\numaptobj}{\numapt_{\mathrm{obj}}}
\newcommand{\numaptill}{\numapt_{\mathrm{ill}}}

\newcommand{\inprod}[2]{\langle#1, #2\rangle}
\newcommand{\conjugate}[1]{\overline{#1}}
\newcommand{\adjoint}[1]{#1^{\ast}}
\newcommand{\Real}{\mathfrak{R}}
\newcommand{\Identity}{\bb{I}}
\newcommand{\Half}{\tfrac{1}{2}}
\newcommand{\argm}{\,\cdot\,}
\newcommand{\CondatVu}{Condat-Vũ}
\newcommand{\elltwo}{\( L^2 \)}
\newcommand{\DataDim}{N}
\newcommand{\DataSpace}{\mathbb{R}^{\DataDim}}

\newcommand{\SigDim}{D}
\newcommand{\SigSpace}{\mathbb{C}^{\SigDim}}
\newcommand{\DiscreteFourier}{\bb{F}}
\newcommand{\ContinuousFourier}{\mathcal{F}}
\newcommand{\USAFTarget}{USAF-1951 phantom}
\newcommand{\darkfield}{dark-field}
\newcommand{\brightfield}{bright-field}

\newcommand{\nonlinear}{nonlinear}
\newcommand{\bb}[1]{\mathbf{#1}}
\newcommand{\cameraman}{\bb{x}^{\mathrm{cm}}}

\DeclarePairedDelimiterX\norm[1]\lVert\rVert{
	\ifblank{#1}{\:\cdot\:}{#1}
}
\DeclarePairedDelimiterX\abs[1]{\lvert}{\rvert}{
	\ifblank{#1}{\:\cdot\:}{#1}
}
\DeclareMathOperator*{\argmin}{arg\,min}
\DeclareMathOperator{\prox}{prox}
\DeclareMathOperator{\proj}{proj}
\definecolor{beamcolor}{RGB}{80, 216, 130}
\definecolor{spycolor}{RGB}{228, 87, 46}
\definecolor{linecolor}{RGB}{82, 113, 166}
\definecolor{darkfieldred}{RGB}{221, 131, 68}
\definecolor{brightfieldyellow}{RGB}{254, 242, 82}
\definecolor{fouriererrorcolor}{RGB}{158, 188, 159}

\title{Perturbative Fourier Ptychographic Microscopy \\for Fast Quantitative Phase Imaging}
\author[1,2]{Martin Zach}
\author[1]{Kuan-Chen Shen}
\author[3]{Ruiming Cao}
\author[1]{Michael Unser}
\author[3]{Laura Waller}
\author[1]{Jonathan Dong\thanks{\texttt{jonathan.dong@epfl.ch}. Martin Zach and Kuan-Chen Shen contributed equally.}}
\affil[1]{Biomedical Imaging Group, École polytechnique fédérale de Lausanne, 1015 Lausanne, Switzerland}
\affil[2]{Center for Biomedical Imaging, 1015 Lausanne, Switzerland}
\affil[3]{Department of Electrical Engineering and Computer Sciences, University of California, Berkeley, CA, 94709, USA}
\begin{document}
\maketitle
\begin{abstract}
    In computational phase imaging with a microscope equipped with an array of light emitting diodes as illumination unit, conventional \glsxtrlong{fpm} achieves high resolution and wide-field reconstructions but is constrained by a lengthy acquisition time.
	Conversely, \gls{dpc} offers fast imaging but is limited in resolution.
	Here, we introduce \gls{pfpm}.
	\gls{pfpm} is an extension of \gls{dpc} that incorporates \darkfield{} illumination to enable fast, high-resolution, wide-field quantitative phase imaging with few measurements.
    We interpret \gls{dpc} as the initial iteration of a Gauss-Newton algorithm with quadratic regularization and generalize it to multiple iterations and more sophisticated regularizers.
	This broader framework is not restricted to \brightfield{} measurements and allows us to overcome resolution limitations of \gls{dpc}.
    We develop tailored annular \darkfield{} illumination patterns that align with the perturbative interpretation and lead to an improvement in the quality of reconstruction with respect to other common illumination schemes.
    Consequently, our methodology combines an enhanced phase reconstruction algorithm with a specialized illumination strategy and offers significant advantages in both imaging speed and resolution.
\end{abstract}
\glsresetall{}
\section{Introduction}
Phase imaging, rooted in the development of the Zernike phase-contrast microscope~\cite{zernike1942phase}, has become an essential technique for the imaging of unstained biological specimens~\cite{park2018quantitative}.
Techniques such as holography~\cite{gabor161new} or Fourier ptychography~\cite{zheng2013wide} retrieve quantitative phase maps via computational algorithms and have found use in many areas of biomedical research, such as histopathology or developmental biology~\cite{park2018quantitative, kim2014high, mir2014label}.

\begin{figure}
	\centering
	\begin{tikzpicture}
		\node at (0, 0) {\includegraphics[width=.6\textwidth]{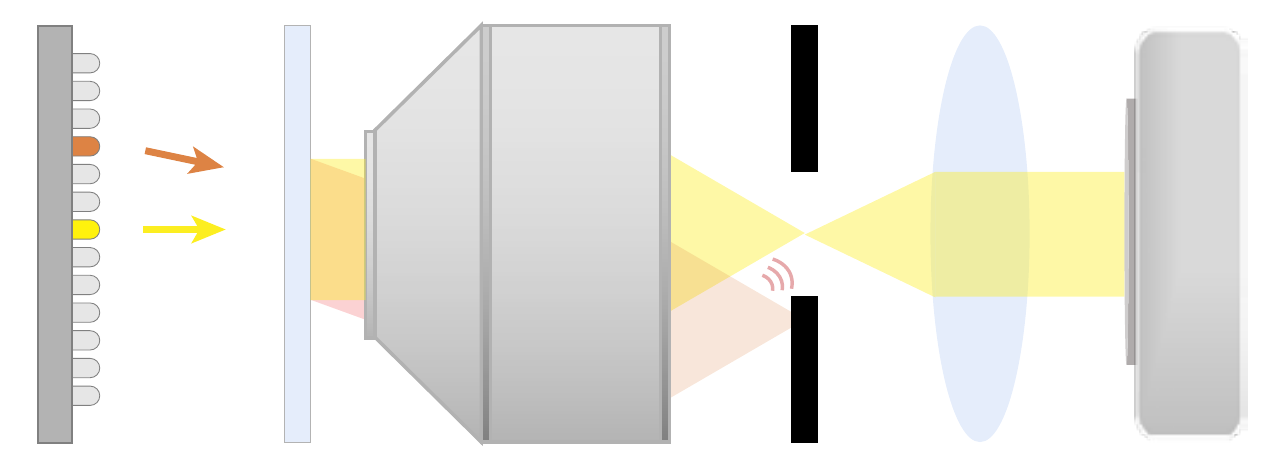}};
		\node (array) at (-6.5, -1.) {\includegraphics[width=2.5cm]{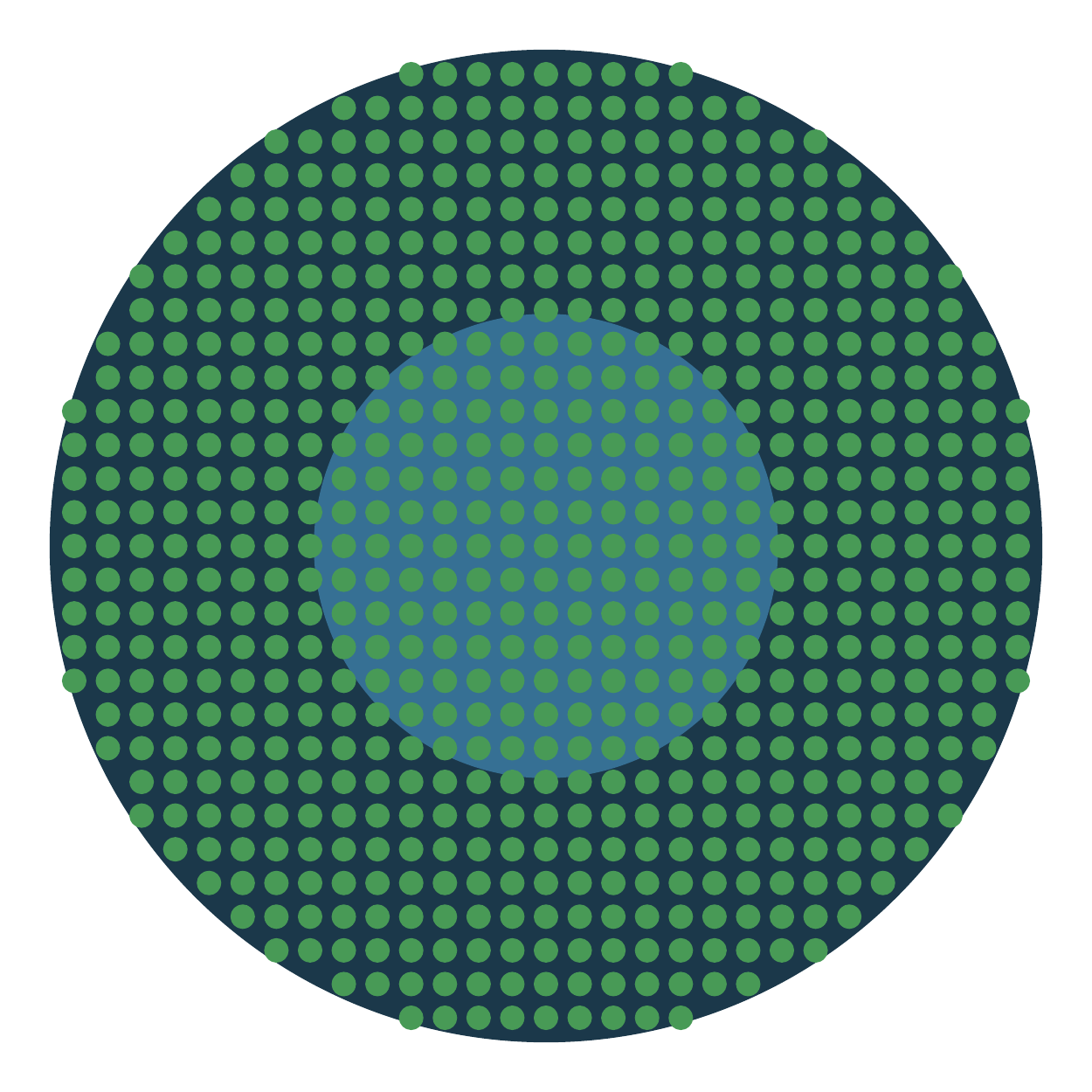}};
		\node [align=center] at (-5.5, -2.5) {Frontal view\\(all \glspl{led} active)};
		\node at (0, -2) {};
		\draw (array) to[out=90, in=180] ++(1.5,1.5);
		\node at (5.8, -1.3) {\includegraphics[cframe=brightfieldyellow 10pt,width=1.5cm]{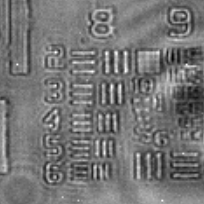}};
		\node [align=center] at (5.7, -2.5) {Measurement from \\\brightfield{} illumination};

		\node at (5.8, 1.3) {\includegraphics[cframe=darkfieldred 10pt,width=1.5cm]{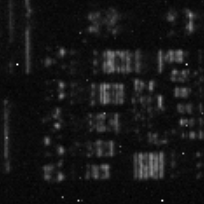}};
		\node [align=center] at (5.7, 2.5) {Measurement from \\\darkfield{} illumination};

		\node at (-.6, 1.8) {Microscope objective};
		\node at (-4.2, 1.8) {LED array};
		\node at (-2.4, 1.8) {Sample};
		\draw (1.2, 1.55) to [in=260,out=45] ++(.4, .25) node [above] {Aperture};
		\node at (4., 1.8) {Camera};
		\node [dashed, draw, rectangle, minimum height=3.3cm, minimum width=5mm] at (1.23, -.04) {};
		\node at (1.15, -1.9) {Fourier space};
		\node [thick, gray, rounded corners, draw, rectangle, minimum width=7.7cm, minimum height=4.5cm] at (.85, 0.) {};
		\node at (.75, -2.5) {Conventional microscope};
	\end{tikzpicture}
	\caption{%
		\gls{led} microscope.
		A programmable \gls{led} array plays the role of the illumination unit.
		In the frontal view of the \gls{led} array on the far left, the inner disk is the \brightfield{} region and the dots are the \glspl{led}.
		The propagation of the illumination from an \gls{led} in the \brightfield{} and the \darkfield{} mode is shown next.
		Corresponding measurements of the \USAFTarget{} are shown on the right.
	}%
	\label{fig:experimental setup}
\end{figure}
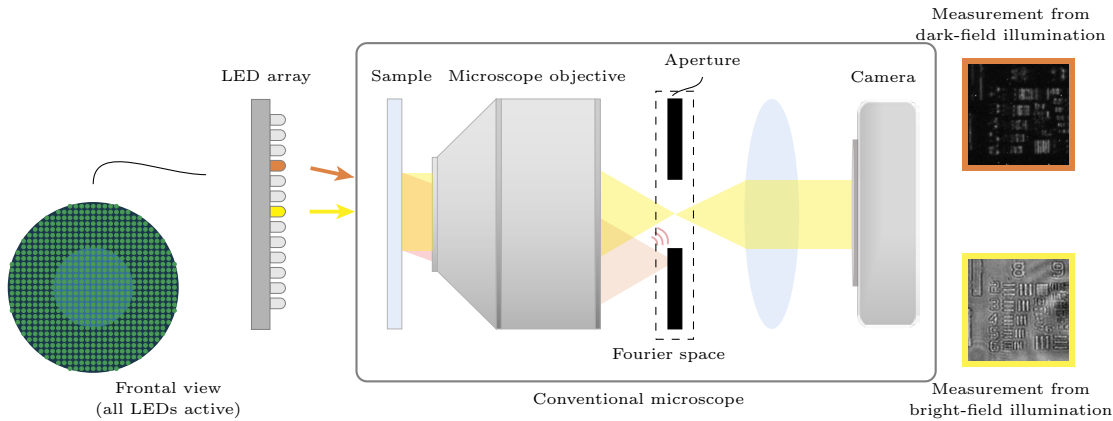

In recent years, the \gls{led}-array microscope has emerged as a versatile tool for computational phase imaging~\cite{zheng2011microscopy, tian2014multiplexed, tian2015quantitative}.
In such a system, users capture measurements under different illumination angles using a programmable illumination unit.
In particular, \glspl{led} positioned within the disc defined by the microscope's \gls{na} provide \brightfield{} illumination, while those that lie outside this disc deliver \darkfield{} illumination, 
as sketched in~\cref{fig:experimental setup}.
Importantly, the illumination patterns are tightly interlinked with the reconstruction algorithm.
Indeed, for \brightfield{}-only illumination, a linear model of the setup is a sufficient approximation when unscattered light is dominant, whereas \darkfield{} illumination necessitates nonlinear recovery algorithms.

The most popular phase imaging methods using \gls{led}-array microscopes are on the two ends of the speed-resolution tradeoff.
On the one hand, \gls{fpm}~\cite{zheng2013wide, zheng2021concept} entails the acquisition of one measurement per (\brightfield{} and \darkfield{}) \gls{led}, typically in the range of hundreds of measurements.
The reconstruction of the high-resolution, wide field-of-view images is done with \nonlinear{} optimization algorithms, typically gradient descent~\cite{yeh2015experimental, dong2023phase}.
On the other hand, \gls{dpc}~\cite{tian2015quantitative, hamilton1984differential, mehta2009quantitative} entails the acquisition of two measurements from asymmetric \brightfield{} illumination patterns.
There, the weak-object approximation enables one to reconstruct the phase by solving a linear system.
However, the linear model only considers \brightfield{} illumination, which limits the achievable resolution.

To image living specimens with acceptable speed, it is essential to minimize the number of required measurements per reconstruction.
\Gls{mfpm}~\cite{tian2014multiplexed} addresses this by the simultaneous use of multiple \glspl{led}, thus merging the speed of \gls{dpc} with the high resolution of \gls{fpm}.
However, the \nonlinear{} optimization becomes increasingly difficult as the number of measurements is reduced, while the optimal design of illumination patterns remains an open problem.
These limitations can partly be addressed by the learning of illumination patterns~\cite{kellman2019data} or the fine tuning of parameters of a reconstruction algorithm~\cite{Xue:19,bohra2023dynamic} which both improve the quality of \gls{mfpm} reconstruction.
However, the learned methods tend to be opaque, hard to interpret, and highly tailored toward specific imaging conditions.

In this work, we introduce \gls{pfpm} for high-resolution phase imaging with few measurements.
We revisit both the reconstruction algorithm and the acquisition strategy from an optimization perspective to tackle the speed-resolution tradeoff.
First, we interpret the \gls{dpc} reconstruction process that is based on the weak-object linearization as the initial iteration of a Gauss-Newton algorithm. 
We then generalize this into a Gauss-Newton framework that accounts for \darkfield{} illumination, with more sophisticated regularization to improve the robustness of phase retrieval.
Second, we design annular \darkfield{} illumination patterns that \enquote{push} reconstructed frequencies into the \darkfield{} region, thereby enhancing resolution.

In summary, our methodology combines a perturbative reconstruction algorithm inspired by \gls{dpc} with tailored illumination patterns to enable fast, high-resolution quantitative phase imaging.
We demonstrate the effectiveness of the proposed approach through numerical simulations and experimental results.

\section{Background}
\subsection{LED-Array Microscope}
The resolution of a classic microscope is determined by the \gls{na} \( \numaptobj \) of its objective, which sets a limit on the largest acquired spatial frequencies.
To bypass this limit, in an \gls{led}-array microscope the conventional illumination unit is replaced by a programmable \gls{led} array.
By illuminating the sample with tilted plane waves, the system gets access to a broader range of spatial frequencies than would be possible under normal incidence.

The illumination \gls{na} \( \numaptill \) is defined as the sine of the largest tilt angle.
When combined with the numerical aperture of the objective, the achievable resolution increases to \( \frac{1.22 \lambda}{\numaptobj + \numaptill} \), where \( \lambda \) is the average illumination wavelength.
Such setups typically use a low-\gls{na} objective to maximize the field-of-view while achieving high resolution in reason of a large illumination \gls{na}.
\Glspl{led} are categorized based on their position relative to the aperture of the microscope:
Bright-field \glspl{led} lie within the cone defined by \( \numaptobj \), thereby allowing unscattered light to pass through the objective and to yield high-intensity measurements that contain low-frequency information. 
In contrast, \darkfield{} \glspl{led} lie outside this cone and generate dim measurements from scattered light that contain high-frequency information.

The model of an \gls{led}-array microscope is sketched in \cref{fig:experimental setup}.
Formally, the illumination of the transmission function \( o: \mathbb{R}^2 \to \mathbb{C} \) of the object with the \( n \)th \gls{led} results in the intensity measurement
\begin{equation}
    y_n(\bb{r}) = \abs{\ContinuousFourier^{-1}\{p \ContinuousFourier \{ o(\argm) \exp(\ImagUnit \inprod{\bb{k}_n}{\argm}) \} \}(\bb{r})}^2,
    \label{eq: sFPM measurement}
\end{equation}
where the \gls{led} index \( n \) ranges from 1 to the total number of \glspl{led} in the programmable illumination unit, denoted \( \Nleds \).
The wave vector \( \bb{k}_n \in \mathbb{R}^{2} \) is associated to the \( n \)th \gls{led} and corresponds to the angle of illumination.
We assume that the microscope is aberration-free and, consequently, characterize it by a pupil function \( p \) that is equal to \( 1 \) inside the circular aperture and \( 0 \) outside.
The symbols \( \mathcal{F} \) and \( \mathcal{F}^{-1} \) denote the two-dimensional Fourier transform and its inverse, respectively, and \( \abs{} \) is the modulus.
We later denote by \( \mathcal{A}_n \{ o \} = \ContinuousFourier^{-1}\{ p \ContinuousFourier \{ o(\argm) \exp{(\ImagUnit \inprod{\bb{k}_n}{\argm})} \} \} \) the field in the camera sensor plane resulting from illumination with the \( n \)th \gls{led}.

The intensity measurements obtained with several \glspl{led} being active are incoherently superimposed.
Formally, the set \( \mathcal{L}_m \subseteq \{ 1, 2, \dotsc, \Nleds \} \) of active \glspl{led}, termed an illumination pattern, results in the intensity measurement
\begin{equation}
    Y_m(\bb{r}) = \sum_{n \in \mathcal{L}_m} y_n(\bb{r}).
    \label{eq:multiplexed measurements}
\end{equation}
The system then ranges through \( m = 1, 2, \ldots, \Npatterns \) distinct illumination patterns to acquire \( \Npatterns \) measurements.
Letting \( Y = (Y_1, Y_2, \dotsc, Y_M) \), we state the relationship between the transmission function \( o \) and the measurements as
\begin{equation}
    Y = G(o),
    \label{eq:acquisition}
\end{equation}
where \( G \) summarizes the nonlinear acquisition process from the \( M \) illumination patterns according to
\begin{equation}
    G(o) = \biggl( \sum_{n \in \mathcal{L}_1} \abs{ \mathcal{A}_n \{ o \}}^2, \dotsc, \sum_{n \in \mathcal{L}_{\Npatterns}} \abs{\mathcal{A}_n \{ o \}}^2 \biggr).
\end{equation}

The challenge in this setup is to find \( M \) illumination patterns and an accompanying algorithm that enables the efficient recovery of the complex-valued transmission function \( o \) from the measurements \( Y \).
Crucially, \( M \) needs to be small to achieve fast imaging and sufficiently large to ensure the robustness of the solution to the nonlinear inverse problem.
In the following sections, we give an overview of proposed illumination patterns and reconstruction algorithms.

\subsection{Conventional Fourier Ptychographic Microscopy}
\begin{figure}
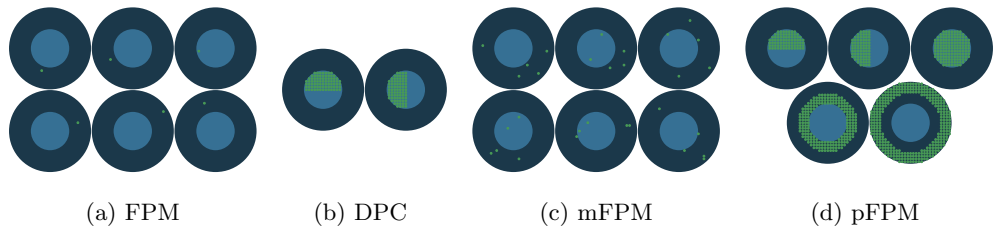

    \centering
	\begin{tikzpicture}[scale=1.1]
        \foreach [count=\iidx from 0] \idx in {111, 212, 301, 433, 544, 600} {
            \pgfmathsetlengthmacro{\myy}{-greater(\iidx, 2)*1cm}
            \pgfmathsetlengthmacro{\myx}{(1*3) * (1cm + .1cm) + mod(\iidx, 3) * 1cm}
            \node at (\myx, \myy) {\includegraphics[width=1.2cm]{results/simulation/generic-patterns/FPM/patterns/\idx.pdf}};
        }
        \pgfmathsetlengthmacro{\myy}{-1cm}
        \pgfmathsetlengthmacro{\myx}{(1*3) * (1cm + .1cm) + mod(5, 3) * 1cm}
        \node at (\myx-1.cm, \myy-1.cm) {\small(a) FPM};
        \foreach \idx in {0, 1} {
            \pgfmathsetlengthmacro{\myy}{-greater(\idx, 2)*1cm -.5cm}
            \pgfmathsetlengthmacro{\myx}{(2*3) * (1cm + .1cm) + mod(\idx, 3) * 1cm}
            \ifthenelse{\idx=0}{\node at (\myx, \myy) {\includegraphics[width=1.2cm]{results/simulation/generic-patterns/DPC/patterns/0\idx.pdf}};}{\node at (\myx, \myy) {\includegraphics[rotate=180,width=1.2cm]{results/simulation/generic-patterns/DPC/patterns/0\idx.pdf}};}
        }
        \node at (\myx+1.75cm, \myy-1.cm) {\small(b) DPC};
        \foreach [count=\iidx from 0] \idx in {1, 2, 3, 4, 5, 6} {
            \pgfmathsetlengthmacro{\myy}{-greater(\iidx, 2)*1cm}
            \pgfmathsetlengthmacro{\myx}{(3*3) * (1cm + .1cm) + mod(\iidx, 3) * 1cm- 1cm}
            \node at (\myx, \myy) {\includegraphics[width=1.2cm]{results/simulation/generic-patterns/mFPM/patterns/0\idx.pdf}};
        }
        \pgfmathsetlengthmacro{\myy}{-greater(5, 2)*1cm}
        \pgfmathsetlengthmacro{\myx}{(3*3) * (1cm + .1cm) + mod(5, 3) * 1cm - 1cm}
        \node at (\myx-1.cm, \myy-1.cm) {\small(c) mFPM};
        \foreach [count=\iidx from 0] \idx in {0, 1, ..., 4} {
            \pgfmathsetlengthmacro{\myy}{-greater(\iidx, 2)*.9cm}
            \pgfmathsetlengthmacro{\myx}{(4*3) * (1cm + .1cm) + mod(\iidx, 3) * 1cm- 1cm + greater(\iidx,2) * .5cm}
            \ifthenelse{\idx=1}{\node at (\myx, \myy) {\includegraphics[width=1.2cm,rotate=180]{results/simulation/generic-patterns/pFPM/patterns/0\idx.pdf}};}{\node at (\myx, \myy) {\includegraphics[width=1.2cm]{results/simulation/generic-patterns/pFPM/patterns/0\idx.pdf}};}
        }
        \node at (\myx+2.25cm, \myy-1.cm) {\small(d) pFPM};
	\end{tikzpicture}
    \caption{%
        Illumination patterns of different methods.
        The inner disk indicates the \brightfield{} region while the dots depict active \glspl{led}.
        The proposed method (\gls{pfpm}) combines \brightfield{} illumination patterns with annular \darkfield{} illumination patterns.
    }\label{fig:illumination patterns}
\end{figure}

Conventional \gls{fpm} dates back to the work of Zheng, Horstmeyer, and Yang~\cite{zheng2013wide}, where single-\gls{led} measurements are employed as illustrated in~\cref{fig:illumination patterns}~(a).
A variety of algorithms address the recovery of the transmission function \( o \) from the intensity measurements;
in the original work, the authors stitch the Fourier space of the reconstruction to be consistent with the intensity measurements in an alternated-projection manner.
Subsequent works optimize the nonlinear least-squares problem associated with~\eqref{eq:acquisition} with gradient-based methods with various acceleration strategies~\cite{yeh2015experimental}.

In the conventional setup, the phase reconstruction requires large overlap ratios of the passband of adjacent \glspl{led} of up to 0.7~\cite{zheng2013wide}.
This necessitates a tight \gls{led} array and, consequently, many (typically a few hundred) measurements.
This leads to long acquisition time and limits the temporal resolution of conventional \gls{fpm}, making it inadequate for fast time-lapse imaging.

\subsection{Multiplexed Fourier Ptychographic Microscopy}\label{sec: mfpm}
To shorten acquisition time, the authors of~\cite{tian2014multiplexed} propose to illuminate the object with multiple, randomly chosen \glspl{led}, thereby reducing the number of measurements needed to achieve results comparable to conventional \gls{fpm}.
Moreover, the increased light throughput from multiple \glspl{led} lowers exposure requirements and therefore enables the imaging of dynamic objects at high frame rates.
Formally, the illumination patterns, sketched in~\cref{fig:illumination patterns} (c), are random subsets of \( \{ 1, 2, \dotsc, \Nleds \} \) of size \( r \) (typically in the single digits), and \( M \) is chosen as \( \Nleds / r \).

Like in conventional \gls{fpm}, researchers often use gradient-based \nonlinear{} optimization algorithms to recover the transmission function \( o \) from the multiplexed intensity measurements \( Y_1, Y_2, \dotsc, Y_{\Npatterns} \)~\cite{tian2014multiplexed, tian2015quantitative}.
More recently, researchers have leveraged machine-learning-based methods to improve the quality of the reconstruction~\cite{zhang2019fourier,Xue:19,wang2024neuph}.

The recourse for as few as possible illumination patterns is paramount to enable the imaging of dynamic objects.
Following the introduction of random patterns in~\cite{tian2014multiplexed}, the authors of~\cite{Tian:15} advocate for a separation of \brightfield{} and \darkfield{} patterns.
They combine the classic \gls{dpc} \brightfield{} patterns with random \darkfield{} patterns, ultimately resulting in \( M = 21 \) measurements.
Xue et al.~\cite{Xue:19} proposed the use of five illumination patterns---two \brightfield{} patterns inspired by \gls{dpc} plus three annular sectors in the \darkfield{}---within a data-driven reconstruction framework.
Instead of fixing them a-priori, Kellmann et al.~\cite{kellman2019physics} learn illumination patterns in a data-driven framework and demonstrate successful reconstructions with a single-digit number of illumination patterns.
The approaches in~\cite{Xue:19} and~\cite{kellman2019physics} ultimately result in reconstruction algorithms that are difficult to interpret and tailored to highly specific imaging conditions or samples. 

\subsection{Differential Phase Contrast}
Instead of relying on nonlinear optimization, \gls{dpc}\footnote{%
    Nowadays, DPC is used in the context of linearized methods with asymmetric illumination patterns, as we use it here.
    Originally, the name stems from the computation of the \enquote{differential} between intensity measurements from two mirrored asymmetric illumination patterns, canceling any influence of absorption~\cite{tian2015quantitative}.
} linearizes the recovery problem around a suitable initial guess and uses direct inversion under suitable illumination patterns to recover an estimation of the transmission function.
For the derivation of the direct inversion, it is convenient to parametrize the transmission function as
\begin{equation}
    o = 1 + \ImagUnit \phi - \mu, 
    \label{eq:object absorption phase}
\end{equation}
where $\phi, \mu: \mathbb{R}^2 \to \mathbb{R}$ are real-valued functions that correspond to phase and absorption, respectively, of the transmission function\footnote{%
    This refers to the linearization of \( (\phi, \mu) \mapsto \exp\bigl( \ImagUnit \phi(\argm) - \mu(\argm) \bigr) \) around \( (0, 0) \).
	This additional linearization can be avoided by the direct parametrization of \( o \) defined in~\eqref{eq:object absorption phase}.
}.
The fundamental assumption in \gls{dpc} is that the object is weak in the sense that both \( \mu \) and \( \phi\) are small.
Then, the nonlinear measurement process~\eqref{eq:acquisition} is well-approximated by the linearization around \( (\mu, \phi) = (0, 0) \) or, equivalently, \( o = \OneFunction \) as
\begin{equation}
    Y \approx G(\OneFunction) + G^\prime(\OneFunction) (o - \OneFunction),
    \label{eq:measurement map}
\end{equation}
where \( G^\prime \) is the Frechet derivative of \( G \), defined at an arbitrary point \( o^k \) by
\begin{equation}
	G^\prime(o^k) : o \mapsto \biggl(
		\sum_{n \in \mathcal{L}_1} 2 \Real (\conjugate{\mathcal{A}_n \{ o^k \}}  \mathcal{A}_n \{ o \}), \dotsc, \sum_{n \in \mathcal{L}_{\Npatterns}} 2\Real(\conjugate{\mathcal{A}_n \{ o^k \}}  \mathcal{A}_n \{ o \}) \biggr).
        \label{eq:frechet continuous}
\end{equation}

Under this approximation, the Fourier transform of the intensity measurement that result from the illumination with the \( n \)th \gls{led} is related to \( \phi \) and \( \mu \) via
\begin{equation}
    \ContinuousFourier \{ y_n \} = \abs{p(\bb{k}_n)}^2 \delta + h^{\mathrm{abs}}_n \ContinuousFourier \{ \mu \} + h^{\mathrm{ph}}_n \ContinuousFourier \{ \phi \},
    \label{eq: DPC measurement}
\end{equation}
where
\begin{equation}
    h^{\mathrm{ph}}_n = \ImagUnit \bigl( \conjugate{p}(\bb{k}_n) p(\bb{k}_n + \argm) - p(\bb{k}_n) \conjugate{p}(\bb{k}_n - \argm) \bigr)
    \label{eq:transfer function phase}
\end{equation}
and
\begin{equation}
	h^{\mathrm{abs}}_n = -\bigl( \conjugate{p}(\bb{k}_n) p(\bb{k}_n + \argm) + p(\bb{k}_n) \conjugate{p}(\bb{k}_n - \argm) \bigr)
    \label{eq:transfer function absorption}
\end{equation}
are the phase- and absorption-transfer function, respectively, where the overbar denotes conjugation and \( \delta \) the Dirac distribution.
A derivation is provided in \cref{sec:derivation}.
Due to the superposition of multiplexed measurements in~\eqref{eq:multiplexed measurements}, they can be accounted for by a summation of the transfer functions for the \glspl{led} of each illumination pattern: \( H^{\mathrm{ph}}_m = \sum_{n \in \mathcal{L}_m} h^{\mathrm{ph}}_n \).
A common simplification~\cite{kellman2019physics} (e.g., in the context of unstained biological samples) involves a phase-only sample, as defined by \( \mu = 0 \) in~\eqref{eq:object absorption phase}.

The presence of noise in the measurements necessitates one to formulate the recovery as a regularized least-squares problem.
For instance, \elltwo{} regularization is used in~\cite{tian2015quantitative}.
Thus, the recovery is posed as the problem
\begin{equation}
    \label{eq:dpc minimization}
	\argmin_{\phi}\,\biggl( \sum_{m=1}^{\Npatterns} \Half \norm{\ContinuousFourier^{-1} \{ H^{\mathrm{ph}}_m \ContinuousFourier \{ \phi \} \} - Y_m}_2^2 + \tfrac{\alpha}{2} \norm{\phi}_2^2 \biggr),
\end{equation}
whose solution is
\begin{equation}
     \ContinuousFourier^{-1}\Biggl\{ 
        \frac{\sum_{m=1}^M \conjugate{H_m^{\mathrm{ph}}}\ContinuousFourier\{Y_m\}}
        {\sum_{m=1}^M \abs[\big]{H_m^{\mathrm{ph}}}^2+\alpha}
    \Biggr\}.
    \label{eq:dpc reconstruction}
\end{equation}
Regarding the design of the illumination patterns, the antisymmetric structure of the phase-transfer function in~\eqref{eq:transfer function phase} suggests the use of antisymmetric \brightfield{} illumination patterns.
To get full coverage of the spectrum, \gls{dpc} is usually done with \( M = 2 \) illumination patterns, where one is rotated by \( \tfrac{\pi}{2} \)~\cite[Fig. 3]{tian2015quantitative},~\cite[Fig. 3 (a)]{kellman2019physics}, see~\cref{fig:illumination patterns} (b).

In summary, \gls{dpc} linearizes the acquisition operator around a suitable initial guess and subsequently estimates the transmission function under some regularization.
This approach is fast and robust, but the linear model requires the weak-object approximation.
Consequently, it is only applicable to measurements from \brightfield{} illumination and limits the achievable resolution improvement to a factor of two.

\section{Method}%
\label{sec:methods}
In this section, we introduce a perturbative formalism to exploit good initial estimates and refine them.
Computationally, we show that the single-shot linearization of \gls{dpc} can be extended to an iterative regularized Gauss-Newton algorithm.
Experimentally, we introduce annular illuminations that take full advantage of this perturbative interpretation to refine the reconstruction using \darkfield{} \glspl{led}.

\subsection{Perturbative \gls{fpm} Algorithm}
We identify \gls{dpc} as the initial iteration of the iterative procedure
\begin{equation}
	o_{k+1} = \argmin_o\,\bigl( \Half \norm{G(o_k) + G^\prime(o_k)(o - o_k) - Y}_2^2 + \alpha R(o) \bigr)
    \label{eq:gn update}
\end{equation}
under the choice \( R(o) = \tfrac{1}{2}\norm{o}_2^2 \) and starting with \( o_0 = \OneFunction \) as the initial guess.
This equivalence is rigorously derived in \cref{sec:todo}.
We recognize~\eqref{eq:gn update} as the proximal Gauss-Newton algorithm~\cite{salzo2012convergence} and generalize \gls{dpc} to more iterations and more sophisticated regularizers.
The generalization to more iterations enables the reconstruction of image features from \darkfield{} measurements, as subsequent iterations of the proximal Gauss-Newton algorithm do not rely on the weak-object approximation.

The generalization to more sophisticated regularizers enables the consideration of regularizers with superior noise-suppression properties.
In this paper, we demonstrate the benefit of this generalization by comparing \gls{tv} regularization to the previously used \elltwo{} regularization.
The framework is not restricted to this; in particular, data-driven regularizers could also be employed.
Indeed, we believe that any phase-retrieval method can benefit from the stronger noise suppression properties of the variational penalties that can be used in the proximal Gauss-Newton framework.

\subsection{Perturbative \gls{fpm} Patterns}
In this section, we detail an acquisition strategy that exploits the combination of \gls{dpc}-type patterns with darkfield annuli to enable fast high-resolution imaging with few illumination patterns that are tailored towards the perturbative approach.
In the \brightfield{}, we exploit the antisymmetric \brightfield{} illumination patterns of \gls{dpc} that enable the accurate recovery of the frequencies of the transmission function up to \( k_{\mathrm{BF}} = 2 \numaptobj / \lambda \) when the object is weak, similar to~\cite{Tian:15}.
In the \darkfield{}, we build on the work of Kellmann et al.\ who show that \darkfield{} illumination patterns need not be antisymmetric---indeed, learned \darkfield{} illumination patterns are actually symmetric~\cite[Fig. 4. (c)]{kellman2019data}, disjoint, and clustered.
Consequently, we partition the \darkfield{} \glspl{led} into annuli, each forming an illumination pattern, to allow for the recovery of higher frequencies from \darkfield{} measurements after the \gls{dpc} solution.
The illumination patterns, shown in~\cref{fig:illumination patterns} (d), enable one to \enquote{push} the recovered frequencies into the \darkfield{}.

Each annulus extends the region of recovered frequencies.
Although the number of annuli is tunable and balances acquisition time with reconstruction quality, our simulations and experiments show excellent results with only two annuli and diminishing returns with more than two annuli, when doubling the brightfield illumination \gls{na}.
Thus, we recommend the use of two \darkfield{} measurements to significantly enhance the quality of the reconstruction at the least expense in imaging speed.

The results in \cref{sec:results} focus on the proposed illumination patterns.
The superiority of the proposed illumination patterns over other popular ones, at a constant number of measurements, is shown in \cref{fig:pattern comparison five} (\( M = 5\)) and \cref{fig:pattern comparison six} (\( M = 6 \)) in \cref{sec:illumination comparison}.
\section{Results}%
\label{sec:results}
In this section, we verify experimentally the proposed reconstruction algorithm and illumination patterns.
In the sequel, \gls{dpc} refers to the direct phase inversion~\eqref{eq:dpc reconstruction} while algorithms with the suffix pFPM use multiple proximal Gauss-Newton iterations.
For the actual practice, we discourage the use of those methods that we have marked with an asterisk (*), even though we report them because they provide additional insight.
BF-pFPM* uses only the two classic \gls{dpc} illumination patterns, BF-pFPM takes advantage of one additional \brightfield{} illumination pattern, and DF-pFPM of two additional \darkfield{} illumination patterns.
For the computational experiments, we use \elltwo{} regularization to match the \gls{dpc} reconstruction.
For the practical experiments, we use \gls{tv} regularization to benefit from its superior noise-suppression properties.
Details such as the number of iterations and the choice of stepsizes are given in~\cref{sec:reconstruction details}.

The results from simulated and experimental data are acquired after a standard discretization of the system on the regular Cartesian grid.
The discretization of the transmission function, the intensity measurements, and all relevant operators, is detailed in~\cref{sec:discrete model}.
We also provide online the code and data to reproduce our results.\footnote{\url{https://github.com/Biomedical-Imaging-Group/perturbative-fpm}}
\subsection{Simulation}
We simulate measurements from a microscope equipped with a \num{10}\(\times\) objective with numerical aperture \( \numaptobj = 0.2 \) and a camera pixel size of \qty{5.5}{\micro\meter}.
An \gls{led} array consisting of (\numproduct{29x29}) \glspl{led} with wavelength \qty{514}{\nano\meter} on a regular grid spaced by \qty{4}{\milli\meter} is positioned \qty{67.5}{\milli\meter} from a phase-only object with phase values in \( [\qty{-0.5}{\radian}, \qty{0.5}{\radian}] \), as derived from the \texttt{cameraman} image provided by \texttt{skimage.data} (see the details in~\cref{sec:simulation data}).
\Glspl{led} up to \( \numaptill = 2 \numaptobj = 0.4 \) are illuminated, which results in a total \gls{na} of \( \numaptobj + \numaptill = 0.6 \).

\begin{figure}
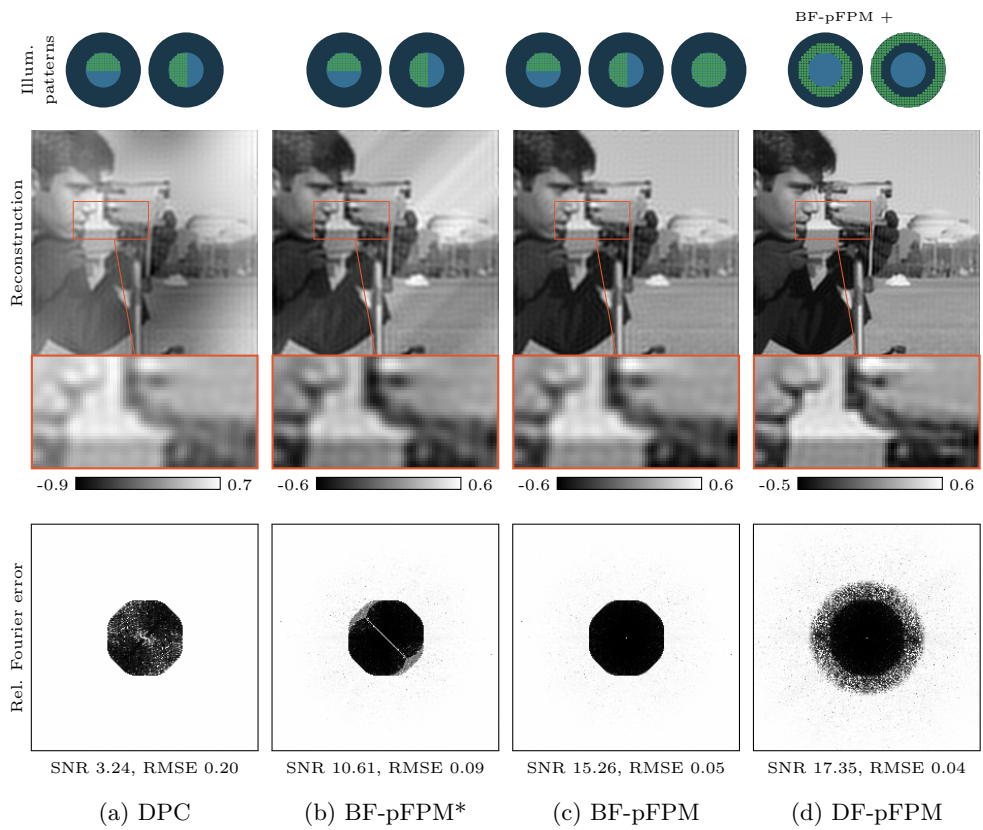

    \centering
	\begin{tikzpicture}
		\pgfmathsetlengthmacro{\wbt}{\imwidth/2}
        \node [rotate=90, align=center] at (1.8, 2.3) {Illum.\\patterns};
        \node [rotate=90] at (1.5, 0) {Reconstruction};
        \node [rotate=90] at (1.5, -5.25) {Rel. Fourier error};
		\foreach \method/\pstart/\pend/\pprev/\anno [count=\imethod] in {DPC/0/1/{}/(a) DPC, BF-pFPMprime/0/1/{}/(b) BF-pFPM*, BF-pFPM/0/2/{}/(c) BF-pFPM, DF-pFPM/3/4/BF-pFPM +/(d) DF-pFPM}
		{
			\pgfmathsetlengthmacro{\myy}{0}
			\pgfmathsetlengthmacro{\myx}{\imethod*(\imwidth + \xpad)}
			\node at (\myx, -7.6cm) {\small\anno};
			\coordinate (onn) at (\myx - .45cm, \myy + 0.3cm);
			\coordinate (att) at (\myx, \myy - 2.25cm);
			\begin{scope}[spy using outlines={rectangle, magnification=3, width=\imwidth, height=\imwidth/2, connect spies}]
				\node at (\myx, \myy) {\includegraphics[width=\imwidth]{./results/simulation/\method/x_est.png}};
				\spy [spycolor] on (onn) in node at (att);
			\end{scope}
			\pgfmathsetlengthmacro{\wbf}{1.1cm}
			\pgfmathsetlengthmacro{\myyy}{\myy + 2.3cm}
			\pgfmathsetlengthmacro{\myyx}{\myx +.5cm - ((\pend - \pstart) * \wbf + .1 cm) / 2 + (- 1) * (\wbf + .1cm)}
			\node at (\myyx+1cm, \myyy+.7cm) {\tiny\pprev};
			\foreach [count=\iipat] \ipat in {\pstart,...,\pend} {
				\pgfmathsetlengthmacro{\myyx}{\myx - ((\pend - \pstart) * \wbf + .0 cm) / 2 + (\iipat - 1) * (\wbf + .0cm)}
				\node at (\myyx, \myyy) {%
					\ifthenelse{\ipat=0}{\includegraphics[width=\wbf]{./results/simulation/\method/patterns/0\ipat.pdf}}{\includegraphics[rotate=180,width=\wbf]{./results/simulation/\method/patterns/0\ipat.pdf}}%
				};
			}
			\node at (\myx, \myy - 3.2cm) {\csvreader[no head]{./results/simulation/\method/span.csv}{}{\csvcoli\drawcolorbar\ \csvcolii}};
			\node [align=left, fill=white, inner sep=0.5mm] at (\myx, \myy - 7cm) {\csvreader[no head]{./results/simulation/\method/metrics.csv}{}{SNR \csvcoli, RMSE \csvcolii}};
			\node at (\myx, \myy - 5.25cm) {\includegraphics[width=\imwidth,cframe=black]{./results/simulation/\method/ft_error.png}};
		}
	\end{tikzpicture}
    \caption{%
		Simulated measurements.
        Top to bottom: llumination patterns; phase reconstruction and magnified region in inset; relative Fourier error~\eqref{eq:relative fourier error}; SNR and RMSE in Fourier space~\eqref{eq:metrics}.
    }%
    \label{fig:simulation result}
\end{figure}

We show the reconstructions from the different methods in \cref{fig:simulation result}.
To facilitate interpretation, we also show the relative Fourier error maps defined by
\begin{equation}
	\abs{\bb{F}\angle(\bb{x}^\ast) - \bb{F} \angle(\hat{\bb{x}})} / \abs{\bb{F}\angle(\bb{x}^\ast)}
    \label{eq:relative fourier error}
\end{equation}
clamped to a maximum of 1, where \( \bb{F} \) is the two-dimensional discrete Fourier transform, \( \bb{x}^* \) the reference object, \( \hat{\bb{x}} \) the reconstruction, and \( \angle \) the element-wise argument of a complex vector.
The relative Fourier error map of BF-pFPM*, shown in \cref{fig:simulation result} (b), demonstrates that multiple Gauss-Newton iterations enable the exact recovery of most frequencies in the passband.
Meanwhile, the \gls{dpc} reconstruction shown in \cref{fig:simulation result} (a) suffers from the errors made in the linearization that corresponds to the weak-object approximation.
However, since pFPM optimizes the complex vector (as opposed to only the phase), ambiguities manifest themselves as large errors along the direction in which the illumination pattern are mutually symmetric.
These errors lead to oriented streaking artifacts in the reconstruction.
The addition of one measurement from a full \brightfield{} illumination mitigates this issue and the algorithm is able to recover frequencies up to \( k_{\mathrm{BF}} \) completely.
Finally, with the addition of measurements from two \darkfield{} illumination patterns with annuli spanning \( [1, 1.5]\numaptobj \) and \( [1.5, 2] \numaptobj \), high frequencies are recovered well.
The qualitative results are supported by the \gls{snr} and \gls{rmse} disclosed in \cref{fig:simulation result}, with
\begin{equation}
	\mathrm{SNR} = 10\log_{10}\frac{\norm{\angle(\bb{x}^*)}^2}{\norm{\angle(\hat{\bb{x}}) - \angle(\bb{x}^*)}^2}
\end{equation}
and
\begin{equation}
	\mathrm{RMSE} = \sqrt{N} \norm{\angle(\hat{\bb{x}}) - \angle(\bb{x}^*)}.
    \label{eq:metrics}
\end{equation} 

\subsection{Experimental}
We image the \USAFTarget{} (a phase-only object manufactured by Benchmark Technologies) on a commercial inverted microscope (Nikon TE2000-U).
The illumination is provided by a customized quasi-dome \gls{led} array (SCI Microscopy)~\cite{phillips2017quasi} with a central wavelength of \qty{525}{\nano\meter} installed on the location of the transmission illumination unit.
The microscope is equipped with a \( 10 \times \) \( 0.25 \) \gls{na} objective lens (Nikon) and captures intensity measurements with a sCMOS sensor (PCO Edge 5.5 monochromatic) that has a pixel size of \qty{6.5}{\micro\meter} using \qty{50}{\milli\second} and \qty{200}{\milli\second} exposure for measurements from \brightfield{} and \darkfield{} illumination patterns, respectively.
We use \glspl{led} up to \( \numaptill = 2 \numaptobj = 0.5 \) for \darkfield{} illumination, which results in a total \gls{na} of \( \numaptobj + \numaptill = 0.75 \).

\begin{figure}
    \centering
    \begin{tikzpicture}
		\pgfmathsetlengthmacro{\myy}{0}
		\foreach \method/\anno [count=\imethod] in {DPC/(a) DPC, BF-pFPMprime/(b) BF-pFPM*, BF-pFPM/(c) BF-pFPM, DF-pFPM/(d) DF-pFPM} {
			\pgfmathsetlengthmacro{\myx}{\imethod*(\imwidth + \xpad)}
			\node at (\myx, -3.7cm) {\small\anno};
			\coordinate (onn) at (\myx + 1.25cm, \myy - 0.4cm);
			\coordinate (att) at (\myx - .75cm, \myy - 2.25cm);
			\begin{scope}[spy using outlines={rectangle, magnification=3, width=\imwidth/2, height=\imwidth/2, connect spies}]
				\node at (\myx, \myy) {\includegraphics[width=\imwidth]{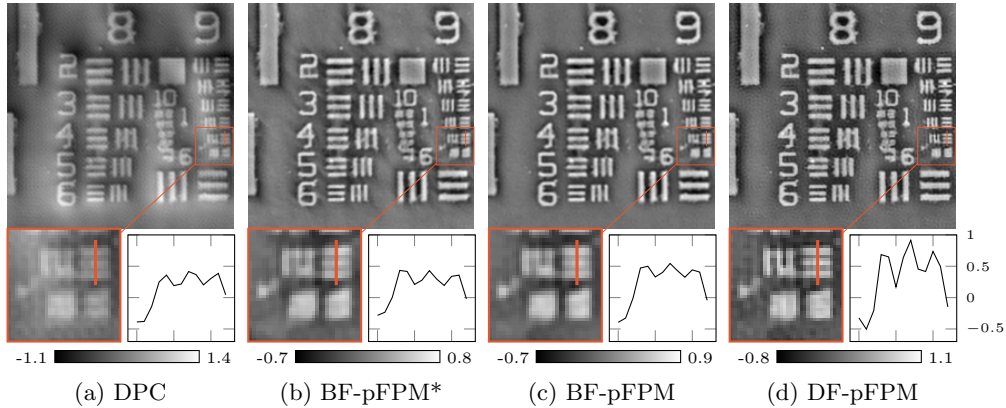}};
				\draw [spycolor] (\myx+1.39cm, \myy-.4cm) -- ++(0cm, .2cm);
				\spy [spycolor] on (onn) in node at (att);
			\end{scope}
			\node at (\myx, \myy - 3.2cm) {\csvreader[no head]{./results/experiments/tv/\method/span.csv}{}{ \csvcoli\drawcolorbar\ \csvcolii}};
			\begin{axis}[width=3cm, height=3cm, at={(\myx + 0.1cm, \myy - 3cm)}, yticklabel pos=right, font=\tiny, ymax=1, ymin=-.7, yticklabel={\ifthenelse{\imethod=4}{\( \pgfmathprintnumber{\tick} \)}{\empty}}, xticklabel=\empty]
				\addplot [mark=none] table [col sep=comma] {./results/experiments/tv/\method/vals.csv};
			\end{axis}
		}
	\end{tikzpicture}
    \caption{%
        Phase reconstruction from measurements of the \USAFTarget.
        From top to bottom: phase reconstruction; magnified region in inset around the fifth and sixth element in the ninth group of the phantom and plot of the line profile of the fifth element in the ninth group.
        The axes of the plot are shared between the methods.
    }%
	\label{fig:experimental results}
\end{figure}

Like in the simulation results, the \gls{dpc} reconstruction suffers from low-frequency artifacts, as shown in \cref{fig:experimental results} (a).
These are removed by BF-pFPM* at the cost of oriented streaking artifacts, as shown in \cref{fig:experimental results} (b).
We conclude from \cref{fig:experimental results} (c) that the additional measurement from full \brightfield{} illumination in BF-pFPM enables satisfactory recovery of all low-frequencies without artifacts.
Finally, the addition of two measurements from \glspl{led} in \darkfield{} annuli spanning \( [1, 1.5] \numaptobj \) and \( [1.5, 2] \numaptobj \) enables the recovery of the high-frequency details.
This is examplified with the profile of the fifth element in Group 9 shown in the insets in \cref{fig:experimental results}:
Unlike the reconstructions from the sole \brightfield{} illumination patterns, DF-pFPM clearly resolved this element and hints at the resolution of the sixth element in Group 9, and even of the first element in Group 10.
Finally, the generalization of the \elltwo{} penalty of \gls{dpc} proves useful, as the DF-pFPM reconstruction is almost free from high-frequency artifacts.
A comparison with the reconstruction obtained by using an \elltwo{} penalty is shown in~\cref{fig:experimental results all} and discussed in~\cref{sec:additional reconstruction results}.

\section{Conclusion}
We introduce a perturbative approach for phase imaging with a \gls{led}-array microscope that enables high-resolution phase imaging with few measurements by combining a perturbative reconstruction algorithm with tailored \brightfield{} and \darkfield{} illumination patterns.
The perturbative algorithm, effectively a Gauss-Newton algorithm, is a generalization of the popular \gls{dpc} to multiple iterations and more sophisticated regularizers with better noise-suppression properties.
We demonstrate the feasibility of the approach on simulated as well as on experimental data.
The proposed method improves the reconstructions significantly over \gls{dpc}, at the cost of only two additional measurements from \darkfield{} illumination and without any learned components.
The proposed illumination patterns also consistently lead to better reconstructions than other illumination patterns that have been proposed previously.
Future work will take advantage of data-driven regularizers to improve reconstructions or decrease exposure time.
In addition, we plan to refine the optical models.
For instance, the present work did not consider aberrations in the pupil function.
Finally, while the current work is dedicated to optical phase imaging with an \gls{led} array, it may be extended to X-ray or electron microscopy.

\section*{Funding}
Jonathan Dong acknowledges funding from the Swiss National Science Foundation (Grant PZ00P2\_216211).
Ruiming Cao was supported in part by Siebel Scholarship.
This work was supported by STROBE:\ A National Science Foundation Science \& Technology Center under Grant No. DMR 1548924. Laura Waller is a Chan Zuckerberg Biohub investigator.

\printbibliography{}

@article{Xue:19,
	author = {Yujia Xue and Shiyi Cheng and Yunzhe Li and Lei Tian},
	journal = {Optica},
	number = {5},
	pages = {618--629},
	publisher = {Optica Publishing Group},
	title = {Reliable deep-learning-based phase imaging with uncertainty quantification},
	volume = {6},
	month = {May},
	year = {2019},
	url = {https://opg.optica.org/optica/abstract.cfm?URI=optica-6-5-618},
	doi = {10.1364/OPTICA.6.000618},
}

@article{zernike1942phase,
	title={Phase contrast, a new method for the microscopic observation of transparent objects},
	author={Zernike, Frits},
	journal={Physica},
	volume={9},
	number={7},
	pages={686--698},
	year={1942},
	publisher={Elsevier}
}

@article{hamilton1984differential,
	title={Differential phase contrast in scanning optical microscopy},
	author={Hamilton, DK and Sheppard, CJR},
	journal={Journal of Microscopy},
	volume={133},
	number={1},
	pages={27--39},
	year={1984},
	publisher={Wiley Online Library}
}

@article{mehta2009quantitative,
	title={Quantitative phase-gradient imaging at high resolution with asymmetric illumination-based differential phase contrast},
	author={Mehta, Shalin B and Sheppard, Colin JR},
	journal={Optics Letters},
	volume={34},
	number={13},
	pages={1924--1926},
	year={2009},
	publisher={Optica Publishing Group}
}

@article{park2018quantitative,
	title={Quantitative phase imaging in biomedicine},
	author={Park, YongKeun and Depeursinge, Christian and Popescu, Gabriel},
	journal={Nature Photonics},
	volume={12},
	number={10},
	pages={578--589},
	year={2018},
	publisher={Nature Publishing Group UK London}
}

@article{gabor161new,
	title={A new microscopic principle},
	author={Gabor, D},
	journal={Nature},
	volume={161},
	pages={777},
	year={1948},
}

@article{zheng2013wide,
	title={Wide-field, high-resolution {Fourier} ptychographic microscopy},
	author={Zheng, Guoan and Horstmeyer, Roarke and Yang, Changhuei},
	journal={Nature Photonics},
	volume={7},
	number={9},
	pages={739--745},
	year={2013},
	publisher={Nature Publishing Group UK London}
}

@article{zheng2011microscopy,
	title={Microscopy refocusing and dark-field imaging by using a simple {LED} array},
	author={Zheng, Guoan and Kolner, Christopher and Yang, Changhuei},
	journal={Optics Letters},
	volume={36},
	number={20},
	pages={3987--3989},
	year={2011},
	publisher={Optica Publishing Group}
}

@article{tian2015quantitative,
	title={Quantitative differential phase contrast imaging in an {LED} array microscope},
	author={Tian, Lei and Waller, Laura},
	journal={Optics Express},
	volume={23},
	number={9},
	pages={11394--11403},
	year={2015},
	publisher={Optica Publishing Group}
}

@article{tian2014multiplexed,
	title={Multiplexed coded illumination for {Fourier} Ptychography with an {LED} array microscope},
	author={Tian, Lei and Li, Xiao and Ramchandran, Kannan and Waller, Laura},
	journal={Biomedical Optics Express},
	volume={5},
	number={7},
	pages={2376--2389},
	year={2014},
	publisher={Optica Publishing Group}
}

@article{bohra2023dynamic,
	title={Dynamic {Fourier} ptychography with deep spatiotemporal priors},
	author={Bohra, Pakshal and Pham, Thanh-an and Long, Yuxuan and Yoo, Jaejun and Unser, Michael},
	journal={Inverse Problems},
	volume={39},
	number={6},
	pages={064005},
	year={2023},
	publisher={IOP Publishing}
}

@article{kellman2019physics,
	title={Physics-based learned design: {Optimized} coded-illumination for quantitative phase imaging},
	author={Kellman, Michael R and Bostan, Emrah and Repina, Nicole A and Waller, Laura},
	journal={IEEE Transactions on Computational Imaging},
	volume={5},
	number={3},
	pages={344--353},
	year={2019},
	publisher={IEEE}
}

@article{kim2014high,
	title={High-resolution three-dimensional imaging of red blood cells parasitized by \textit{{Plasmodium} falciparum} and \textit{in situ} hemozoin crystals using optical diffraction tomography},
	author={Kim, Kyoohyun and Yoon, HyeOk and Diez-Silva, Monica and Dao, Ming and Dasari, Ramachandra R and Park, YongKeun},
	journal={Journal of Biomedical Optics},
	volume={19},
	number={1},
	pages={011005--011005},
	year={2014},
	publisher={Society of Photo-Optical Instrumentation Engineers}
}

@article{mir2014label,
	title={Label-free characterization of emerging human neuronal networks},
	author={Mir, Mustafa and Kim, Taewoo and Majumder, Anirban and Xiang, Mike and Wang, Ru and Liu, S Chris and Gillette, Martha U and Stice, Steven and Popescu, Gabriel},
	journal={Scientific Reports},
	volume={4},
	number={1},
	pages={4434},
	year={2014},
	publisher={Nature Publishing Group UK London}
}

@inproceedings{phillips2017quasi,
	title={Quasi-dome: {A} self-calibrated high-{NA} {LED} illuminator for {Fourier} ptychography},
	author={Phillips, Zachary F and Eckert, Regina and Waller, Laura},
	booktitle={Imaging Systems and Applications},
	pages={IW4E--5},
	year={2017},
}

@article{salzo2012convergence,
	title = {Convergence analysis of a proximal {Gauss}-{Newton} method},
	volume = {53},
	ISSN = {1573-2894},
	url = {http://dx.doi.org/10.1007/s10589-012-9476-9},
	DOI = {10.1007/s10589-012-9476-9},
	number = {2},
	journal = {Computational Optimization and Applications},
	publisher = {Springer Science and Business Media LLC},
	author = {Salzo,  Saverio and Villa,  Silvia},
	year = {2012},
	month = mar,
	pages = {557–589}
}

@article{chambolle2015ergodic,
	title = {On the ergodic convergence rates of a first-order primal-dual algorithm},
	volume = {159},
	ISSN = {1436-4646},
	url = {http://dx.doi.org/10.1007/s10107-015-0957-3},
	DOI = {10.1007/s10107-015-0957-3},
	number = {1–2},
	journal = {Mathematical Programming},
	publisher = {Springer Science and Business Media LLC},
	author = {Chambolle,  Antonin and Pock,  Thomas},
	year = {2015},
	month = oct,
	pages = {253–287}
}

@article{chambolle2004algorithm,
	author = {Chambolle, Antonin},
	date = {2004/01/01},
	doi = {10.1023/B:JMIV.0000011325.36760.1e},
	id = {Chambolle2004},
	isbn = {1573-7683},
	journal = {Journal of Mathematical Imaging and Vision},
	number = {1},
	pages = {89--97},
	title = {An Algorithm for Total Variation Minimization and Applications},
	url = {https://doi.org/10.1023/B:JMIV.0000011325.36760.1e},
	volume = {20},
	year = {2004}
}

@InProceedings{porta2024inexact,
	author={Porta, Federica and Villa, Silvia and Viola, Marco and Zach, Martin},
	editor={Benfenati, Alessandro and Porta, Federica and Bubba, Tatiana Alessandra and Viola, Marco},
	title={On the Inexact Proximal {Gauss}-{Newton} Methods for Regularized Nonlinear Least Squares Problems},
	booktitle={Advanced Techniques in Optimization for Machine Learning and Imaging},
	year={2024},
	publisher={Springer Nature Singapore},
	address={Singapore},
	pages={151--165},
	isbn={978-981-97-6769-4}
}

@inproceedings{kellman2019data,
	title={Data-driven design for {Fourier} ptychographic microscopy},
	author={Kellman, Michael and Bostan, Emrah and Chen, Michael and Waller, Laura},
	booktitle={2019 IEEE International Conference on Computational Photography (ICCP)},
	pages={1--8},
	year={2019},
    address={Tokyo, Japan}
}

@article{dong2023phase,
	title={Phase retrieval: From computational imaging to machine learning: A tutorial},
	author={Dong, Jonathan and Valzania, Lorenzo and Maillard, Antoine and Pham, Thanh-an and Gigan, Sylvain and Unser, Michael},
	journal={IEEE Signal Processing Magazine},
	volume={40},
	number={1},
	pages={45--57},
	year={2023},
	publisher={IEEE}
}

@article{yeh2015experimental,
	title={Experimental robustness of {Fourier} ptychography phase retrieval algorithms},
	author={Yeh, Li-Hao and Dong, Jonathan and Zhong, Jingshan and Tian, Lei and Chen, Michael and Tang, Gongguo and Soltanolkotabi, Mahdi and Waller, Laura},
	journal={Optics Express},
	volume={23},
	number={26},
	pages={33214--33240},
	year={2015},
	publisher={Optica Publishing Group}
}

@article{zhang2019fourier,
	title={Fourier ptychographic microscopy reconstruction with multiscale deep residual network},
	author={Zhang, Jizhou and Xu, Tingfa and Shen, Ziyi and Qiao, Yifan and Zhang, Yizhou},
	journal={Optics Express},
	volume={27},
	number={6},
	pages={8612--8625},
	year={2019},
	publisher={Optica Publishing Group}
}

@article{zheng2021concept,
	title={Concept, implementations and applications of {Fourier} ptychography},
	author={Zheng, Guoan and Shen, Cheng and Jiang, Shaowei and Song, Pengming and Yang, Changhuei},
	journal={Nature Reviews Physics},
	volume={3},
	number={3},
	pages={207--223},
	year={2021},
	publisher={Nature Publishing Group UK London}
}

@article{wang2024neuph,
	title={NeuPh: {Scalable} and generalizable neural phase retrieval with local conditional neural fields},
	author={Wang, Hao and Zhu, Jiabei and Li, Yunzhe and Yang, Qianwan and Tian, Lei},
	journal={Advanced Photonics Nexus},
	volume={3},
	number={5},
	pages={056005--056005},
	year={2024},
	publisher={Society of Photo-Optical Instrumentation Engineers}
}

@article{Tian:15,
	author = {Lei Tian and Ziji Liu and Li-Hao Yeh and Michael Chen and Jingshan Zhong and Laura Waller},
	journal = {Optica},
	number = {10},
	pages = {904--911},
	publisher = {Optica Publishing Group},
	title = {Computational illumination for high-speed in vitro {Fourier} ptychographic microscopy},
	volume = {2},
	month = {Oct},
	year = {2015},
	url = {https://opg.optica.org/optica/abstract.cfm?URI=optica-2-10-904},
	doi = {10.1364/OPTICA.2.000904},
}

\appendix
\section{Derivation of the Transfer Functions}
\label{sec:derivation}
After parametrizing the transmission function \( o \) by \( 1 + \ImagUnit \phi - \mu\), the nonlinear measurement operator of the \( n \)th \gls{led} is
\begin{equation}
        (\phi, \mu) \mapsto y_n(\bb{r}) = \abs{\ContinuousFourier^{-1} \{p\ContinuousFourier\{ \bigl( 1 + \ImagUnit \phi - \mu \bigr) \exp(\ImagUnit \langle \bb{k}_n, \argm\rangle) \} \}(\bb{r})}^2.
\end{equation}
Using the shorthand \( \FT{f} \coloneqq \ContinuousFourier\{f\} \) and the identities \( \ContinuousFourier \{ 1 \} = \delta \), \( \ContinuousFourier \{ f \exp(\ImagUnit \inprod{\bb{k}_n}{\argm}) \} = \FT{f}(\argm - \bb{k}_n) \), the right hand side becomes\footnote{%
    We give \enquote{evaluation} higher precedence than \enquote{multiplication,} so that \( p \bigl( \delta + \ImagUnit \FT{\phi} - \FT{\mu} \bigr) (\argm - \bb{k}_n) \) is the map \( \bb{k} \mapsto  p(\bb{k})\delta(\bb{k} - \bb{k}_n) + \ImagUnit p(\bb{k})\FT{\phi}(\bb{k} - \bb{k}_n) - p(\bb{k})\FT{\mu}(\bb{k} - \bb{k}_n) \).
}
\begin{equation}
    y_n(\bb{r}) = \abs{\ContinuousFourier^{-1} \{p \bigl( \delta + \ImagUnit \FT{\phi} - \FT{\mu} \bigr) (\argm - \bb{k}_n) \}(\bb{r})}^2
	\label{eq:intermediate}
\end{equation}
From \( \abs{z}^2 = z\conjugate{z}\), we rewrite \eqref{eq:intermediate} as
\begin{equation}
    y_n(\bb{r}) = \bigl( \ContinuousFourier^{-1} \{p \bigl( \delta + \ImagUnit \FT{\phi} - \FT{\mu} \bigr) (\argm - \bb{k}_n) \}(\bb{r})\bigr) \bigl( \conjugate{\ContinuousFourier^{-1} \{p \bigl( \delta + \ImagUnit \FT{\phi} - \FT{\mu} \bigr) (\argm - \bb{k}_n) \}(\bb{r})} \bigr).
    \label{eq:product}
\end{equation}
Since \( \conjugate{\ContinuousFourier^{-1} \{ \FT{f} \}} = \ContinuousFourier^{-1} \left\{ \conjugate{\FT{f}(- \argm)} \right\} \), we write the second factor of \eqref{eq:product} as
\begin{equation}
\begin{aligned}
       \conjugate{\ContinuousFourier^{-1} \{p \bigl( \delta + \ImagUnit \FT{\phi} - \FT{\mu} \bigr) (\argm - \bb{k}_n) \}} &= \ContinuousFourier^{-1} \{\conjugate{p}(-\argm) \conjugate{\bigl( \delta + \ImagUnit \FT{\phi} - \FT{\mu} \bigr) (-\argm - \bb{k}_n)} \}\\ 
        &= \ContinuousFourier^{-1} \{\conjugate{p}(-\argm) \bigl( \delta(\argm + \bb{k}_n) - (\ImagUnit \FT{\phi} + \FT{\mu})(\argm + \bb{k}_n)  \bigr)  \}.
\end{aligned}
\end{equation}
Now, we compute the product in~\eqref{eq:product}.
We use that \( \ContinuousFourier^{-1} \{ p\delta(\argm - \bb{k}_n) \} \ContinuousFourier^{-1} \{ \FT{f} \} = p(\bb{k}_n) \ContinuousFourier^{-1} \{ \FT{f} (\argm - \bb{k}_n) \} \) and keep only terms up to first order in \( \FT{\mu} \) and \( \FT{\phi} \) to establish that~\eqref{eq:product} can be expressed as
\begin{equation}
    y_n(\bb{r}) = \abs{p(\bb{k}_n)}^2 + \conjugate{p(\bb{k}_n)} \ContinuousFourier^{-1} \{ p(\bb{k}_n + \argm)(\ImagUnit \FT{\phi} - \FT{\mu}) \}(\bb{r}) - p(\bb{k}_n) \ContinuousFourier^{-1} \{ \conjugate{p(\bb{k}_n - \argm)} (j \FT{\phi} + \FT{\mu})\}(\bb{r}).
    \label{eq:dpc intermediate result}
\end{equation}
Taking the Fourier transform of~\eqref{eq:dpc intermediate result} and grouping terms in \( \FT{\mu} \) and \( \FT{\phi} \), we obtain that that 
\begin{equation}
    \begin{aligned}
        \FT{y}_n = {} & \abs{p(\bb{k}_n)}^2 \delta \\
        &+ \FT{\phi} \ImagUnit \bigl( \conjugate{p(\bb{k}_n)} p(\bb{k}_n + \argm) - p(\bb{k}_n)\conjugate{p(\bb{k}_n - \argm)} \bigr) \\&- \FT{\mu} \bigl( \conjugate{p(\bb{k}_n)} p(\bb{k}_n + \argm) + p(\bb{k}_n)\conjugate{p(\bb{k}_n - \argm)} \bigr),
     \end{aligned}
\end{equation}
and the transfer functions for \( \phi \) and \( \mu \) in \eqref{eq:transfer function phase} and \eqref{eq:transfer function absorption} can be read of as the functions multiplying \( \FT{\phi} \) and \( \FT{\mu} \), respectively.
\section{The First Iteration of Proximal Gauss-Newton Is DPC}
\label{sec:todo} 
The initial Gauss-Newton step, where the measurement operator is linearized around \( o = 1 \), solves the problem
\begin{equation}
	\argmin_{o}\, \Bigl( \Half \norm{G(1) + G^\prime(1)(o - 1) - Y}_2^2 + \tfrac{\alpha}{2} \norm{o}_2^2 \Bigr).
\end{equation}
First, we note that \( G(\OneFunction) = \bigl(\sum_{n \in \mathcal{L}_1} \abs{p(\bb{k}_n)}^2, \dotsc, \sum_{n \in \mathcal{L}_M} \abs{p(\bb{k}_n)}^2 \bigr) \).
Since \( (o - \OneFunction) = (\ImagUnit \phi - \mu) \), we have that
\begin{equation}
    G^\prime(1)(o - 1) = \biggl( \sum_{n \in \mathcal{L}_1} 2\Real (\conjugate{\mathcal{A}_n\{ 1 \}}\mathcal{A}_n\{ \ImagUnit \phi - \mu \} ), \dotsc, \sum_{n \in \mathcal{L}_M} 2 \Real ( \conjugate{\mathcal{A}_n \{ 1 \}}\mathcal{A}_n\{ \ImagUnit \phi - \mu \} ) \biggr).
\end{equation}
Thus, it suffices to analyze \( 2 \Real( \conjugate{\mathcal{A}_n\{ 1 \}}\mathcal{A}_n\{ j\phi - \mu \}) \) for an arbitrary \( n \in \{ 1, 2, \dotsc, \Nleds \} \).
Using \( \conjugate{\mathcal{A}_n \{ 1 \}} = \conjugate{p(\bb{k}_n)}\exp(\ImagUnit \inprod{-\bb{k}_n}{\argm}) \) and inserting the definition of \( \mathcal{A}_n \), we obtain that
\begin{equation}
    \begin{aligned}
\MoveEqLeft
        2 \Real ( \conjugate{\mathcal{A}_n\{ 1 \}}\mathcal{A}_n\{ \ImagUnit \phi - \mu \} ) \\
		&=2 \Real ( \conjugate{p(\bb{k}_n)}\exp(\ImagUnit \inprod{-\bb{k}_n}{\argm}) \ContinuousFourier^{-1}\{ p \ContinuousFourier \{ (\ImagUnit  \phi(\argm) - \mu(\argm)) \exp{(\ImagUnit \inprod{\bb{k}_n}{\argm})} \} \} ) \\
        &=2 \Real ( \conjugate{p(\bb{k}_n)}\exp(\ImagUnit \inprod{-\bb{k}_n}{\argm}) \ContinuousFourier^{-1}\{ p (\ImagUnit \FT{\phi}(\argm - \bb{k}_n) - \FT{\mu}(\argm - \bb{k}_n)) \} ) \\
        &=2 \Real ( \conjugate{p(\bb{k}_n)} \ContinuousFourier^{-1}\{p(\argm + \bb{k}_n)(\ImagUnit \FT{\phi} - \FT{\mu}) \} ) \\
		&=\conjugate{p(\bb{k}_n)} \ContinuousFourier^{-1}\{p(\argm + \bb{k}_n) (\ImagUnit \FT{\phi} - \FT{\mu} ) \} + \conjugate{\conjugate{p(\bb{k}_n)} \ContinuousFourier^{-1}\{p(\argm + \bb{k}_n)( \ImagUnit \FT{\phi} - \FT{\mu}  \})},
    \end{aligned}
\end{equation}
where the last equality is due to \( 2 \Real ( f ) = f + \conjugate{f}\).
Now, we simplify the second term using \( \conjugate{\ContinuousFourier^{-1} \{ \FT{f} \}} = \ContinuousFourier^{-1} \{ \conjugate{\FT{f}(- \argm)} \} \) and \( \conjugate{\FT{f}} = \FT{f}(-\argm) \) for real \( f \), which leads to
\begin{equation}
    \begin{aligned}
        \conjugate{\conjugate{p(\bb{k}_n)} \ContinuousFourier^{-1}\{p(\argm + \bb{k}_n) (\ImagUnit \FT{\phi} - \FT{\mu}) \}}
        &= p(\bb{k}_n) \ContinuousFourier^{-1}\{ \conjugate{p(\bb{k}_n - \argm) ( \ImagUnit \FT{\phi}(-\argm) - \FT{\mu}(-\argm)) } \} \\
        &=-p(\bb{k}_n) \ContinuousFourier^{-1}\{ \conjugate{p(\bb{k}_n - \argm)} (\ImagUnit  \FT{\phi} + \FT{\mu}) \}.
    \end{aligned}
\end{equation}
Thus, we conclude that
\begin{equation}
        2 \Real ( \conjugate{\mathcal{A}_n\{ 1 \}}\mathcal{A}_n\{ \ImagUnit \phi - \mu \} ) = 
       \conjugate{p(\bb{k}_n)} \ContinuousFourier^{-1}\{p(\argm + \bb{k}_n) (\ImagUnit \FT{\phi} - \FT{\mu} ) \} -p(\bb{k}_n) \ContinuousFourier^{-1}\{ \conjugate{p(\bb{k}_n - \argm)} (\ImagUnit \FT{\phi} + \FT{\mu}) \}.
\end{equation}
This expression is the same as \eqref{eq:dpc intermediate result}, except for the missing term \( \abs{p(\bb{k}_n)}^2 \) which is due to the contribution of the \( n \)th \gls{led} in \( G(1) \).

This proves that the first iteration of a proximal Gauss-Newton method corresponds to \gls{dpc}, as the linearized model in our Gauss-Newton algorithm matches the conventional \gls{dpc} transfer function and since both minimize the same \elltwo{} loss, assuming that $\phi$ is zero-mean.

\section{Discretization}%
\label{sec:discrete model}
We discretize on the two-dimensional uniform Cartesian grid both the object space in \( \{ 1, 2, \dotsc, d_1 \} \times \{ 1, 2, \dotsc, d_2 \} \) and the image space in \( \{ 1, 2, \dotsc, n_1 \} \times \{ 1, 2, \dotsc, n_2 \} \). 
The discrete phase image $\bb{x}^\ast \in \SigSpace$ is of dimension $\SigDim = d_1 d_2$ and the output intensities $\bb{y}_n$ are in $\mathbb{R}^\DataDim$ with $\DataDim = n_1 n_2$. 
In the discrete setting, the intensity measurements~\eqref{eq: sFPM measurement} are modeled as
\begin{equation}
	\bb{y}_n = \abs{\bb{A}_n \bb{x}^\ast}^2
\end{equation}
where the linear operator corresponding to the \( n \)th \gls{led} is defined as $\bb{A}_n = \adjoint{\DiscreteFourier} \bb{C} \bb{P} \bb{S}_n \DiscreteFourier$, with \( \DiscreteFourier \in \mathbb{C}^{D \times D} \) the two-dimensional discrete Fourier transform, \( \bb{S}_n \in \mathbb{C}^{D \times D} \) a circular shift in Fourier space, \( \bb{C} \in \mathbb{C}^{D \times N} \) a crop to reduce output dimension, \(\bb{P} \in \mathbb{C}^{N \times N} \) a model of the pupil function of the microscope, and finally \( \adjoint{\DiscreteFourier} \in \mathbb{C}^{N \times N} \) the two-dimensional inverse discrete Fourier transform. 
The phase modulation by a circular shift in Fourier space is represented by \( \bb{S}_n \).
Moreover, we round the theoretical shift determined by the \gls{led} position and the optical system to the nearest integer to avoid interpolation.
Since we assume an aberration-free system, \( \bb{P} \) acts as a binary diagonal operator that blanks frequencies outside of the aperture.

For multiplexed measurements, let \( \mathcal{L}_1, \mathcal{L}_2, \dotsc, \mathcal{L}_{\Npatterns} \) be sets of \gls{led} indices that encode the \( \Npatterns \) illumination patterns, as in \cref{sec: mfpm}.
For each \( m = 1, 2, \dotsc, \Npatterns, \) the measurements are
\begin{equation}
	\bb{Y}_m = \sum_{n \in \mathcal{L}_m} \abs{\bb{A}_n \bb{x}^\ast}^2,
\end{equation}
or, after letting \( \bb{Y} = (\bb{Y}_1, \bb{Y}_2, \dotsc, \bb{Y}_{\Npatterns}) \),
\begin{equation}
	\bb{Y} = \mathrm{G}(\bb{x}^\ast),
	\label{eq:data}
\end{equation}
where
\begin{equation}
    \mathrm{G} : \SigSpace \to {(\DataSpace)^{\Npatterns}} : \bb{x} \mapsto \biggl(
		\sum_{n \in \mathcal{L}_1} \abs{\bb{A}_n \bb{x}}^2,
			\dotsc,
		\sum_{n \in \mathcal{L}_{\Npatterns}} \abs{\bb{A}_n \bb{x}}^2 \biggr),
\end{equation}
summarizes the \nonlinear{} multiplexed measurement acquisition.
The Fr\'echet derivative of \( G \) at some point \( \bb{x}^k \) is the map
\begin{equation}
	\mathrm{G}^\prime(\bb{x}^k) : \SigSpace \to {(\DataSpace)^{\Npatterns}} : \bb{x} \mapsto \biggl(
		\sum_{n \in \mathcal{L}_1} \Real (2 \conjugate{\bb{A}_n \bb{x}^k} \odot \bb{A}_n \bb{x} ),
			\dotsc,
		\sum_{n \in \mathcal{L}_{\Npatterns}} \Real ( 2 \conjugate{\bb{A}_n \bb{x}^k} \odot \bb{A}_n \bb{x} ) \biggr),
\end{equation}
where $\odot$ denotes an element-wise multiplication.

\section{Details of the Reconstruction Algorithm}%
\label{sec:reconstruction details}
In our setup, the regularized nonlinear least-squares problem associated with~\eqref{eq:data} is
\begin{equation}
	\argmin_{\bb{x}}\,\bigl(\Half \norm{\mathrm{G}(\bb{x}) - \bb{Y}}_2^2 + \alpha R(\bb{x})\bigr).
    \label{eq:regularized least squares}
\end{equation}
Given the current solution \( \bb{x}_k \), a proximal Gauss-Newton iteration on~\eqref{eq:regularized least squares} amounts to an update rule of the form
\begin{equation}
	\bb{x}_{k + 1} = \prox_{\alpha R}^{H(\bb{x}_k)} \Bigl( \bb{x}_k - \bigl( H(\bb{x}_k) \bigr)^{-1}\adjoint{\bigl( \mathrm{G}^\prime(\bb{x}_k) \bigr)} \bigl( \mathrm{G}(\bb{x}_k) - \bb{Y} \bigr) \Bigr),
	\label{eq:pgn}
\end{equation}
where the asterisk denotes the adjoint and \( \prox_{\alpha R}^{H(\bb{x}_k)} \) is the proximal operator of \( \alpha R \) with respect to the metric \( H(\bb{x}_k) = \adjoint{\bigl( G^\prime(\bb{x}_k) \bigr)}G^\prime(\bb{x}_k) \).
The proximal operator is defined as
\begin{equation}
	\prox_{\alpha R}^{H(\bb{x}_k)}(\bb{x}) = \argmin_{\bb{z}}\, \bigl( \Half \norm{\bb{z} - \bb{x}}^2_{H(\bb{x}_k)} + \alpha R(\bb{z}) \bigr).
\end{equation}
Proposition 6 in \cite{salzo2012convergence} states that~\eqref{eq:pgn} is equivalent to the update rule
\begin{equation}
	\bb{x}_{k+1} = \argmin_{\bb{x}}\, \bigl(\Half \norm{\mathrm{G}(\bb{x}_k) + \mathrm{G}^\prime(\bb{x}_k)(\bb{x} - \bb{x}_k) - \bb{Y}}_2^2 + \alpha R(\bb{x}) \bigr).
	\label{eq:pgn subproblem}
\end{equation}
Indeed, the proximal Gauss-Newton algorithm can be implemented by iteratively solving the regularized linearized problem.

We consider two regularizers: Classical \elltwo{} regularization \( \bb{x} \mapsto \Half \norm{\bb{x}}_2^2 \), and isotropic \gls{tv} regularization \( \bb{x} \mapsto \norm{\FFDop \bb{x}}_{2, 1} \), where \( \FFDop : \SigSpace \to \mathbb{C}^{\SigDim \times 2}\) is a forward finite-difference operator with Neumann boundaries and \( \norm{}_{2, 1} : \mathbb{C}^{\SigDim \times 2} \to \mathbb{R}_+ : \bb{x} \mapsto \sum_{d=1}^{\SigDim}\sqrt{\abs{\bb{x}_{d, 1}}^2 + \abs{\bb{x}_{d, 2}}^2} \).

For \elltwo{}, we solve the linear system that results from the first-order optimality condition
\begin{equation}
	\Bigl( \adjoint{\bigl( \mathrm{G}^\prime(\bb{x}_k) \bigr)} \mathrm{G}^\prime(\bb{x}_k) + \alpha \Identity \Bigr) \bb{x} = \adjoint{\bigl( \mathrm{G}^\prime(\bb{x}_k) \bigr)} \bigl( \mathrm{G}^\prime(\bb{x}_k)\bb{x}_k + \bb{Y} - \mathrm{G}(\bb{x}_k) \bigr),
\end{equation}
where \( \Identity \) is the identity matrix, using the conjugate gradient algorithm.
This also covers the unregularized case \( \alpha = 0 \).
For \gls{tv}, we use the dual representation of the \gls{tv} norm, given by
\begin{equation}
	\alpha \norm{\FFDop\,\cdot\,}_{2, 1} = \max_{\bb{z} \in \mathbb{C}^{D \times 2}}\ \bigl( \langle \argm, \adjoint{\FFDop} \bb{z} \rangle - \delta_{\norm{}_{2,\infty} \leq \alpha}(\bb{z}) \bigr),
\end{equation}
where
\begin{equation}
    \delta_{\norm{}_{2,\infty} \leq \alpha}(\bb{z}) = \begin{cases}
        \infty, &\ \text{if}\ \max_{d \in \{ 1, 2, \dotsc, D \}} \norm{\bb{z}_d}_{2} > \alpha \\ 
        0, &\ \text{else}.
    \end{cases}
\end{equation}
We then solve the nonsmooth convex-concave saddle point problem
\begin{equation}
	\argmin_{\bb{x} \in \mathbb{C}^D} \max_{\bb{z} \in \mathbb{C}^{D \times 2}} \bigl( \Half \norm{\mathrm{G}(\bb{x}_k) + \mathrm{G}^\prime(\bb{x}_k)(\bb{x} - \bb{x}_k) - \bb{Y}}_2^2 + \langle \bb{x}, \adjoint{\FFDop} \bb{z} \rangle - \delta_{\norm{}_{2,\infty} \leq \alpha}(\bb{z}) \bigr)
	\label{eq:dual subproblem}
\end{equation}
using the \CondatVu{} algorithm, an extension of the primal-dual hybrid gradient algorithm with explicit gradient steps, as summarized in~\cref{alg:cv}.
There,
\begin{equation}
	\bb{z} = \proj_{\{ \norm{}_{2, \infty} < \alpha \}}(\bar{\bb{z}}) \iff \bb{z}_{d} = \frac{\bar{\bb{z}}_{d}}{\max(1, \alpha^{-1} \norm{\bar{\bb{z}}_{d}}_2)}
\end{equation}
is a pointwise projection onto the two-norm ball and at the \( k \)th \gls{pgn} iteration, and we chose the step sizes \( \tau, \sigma > 0 \) as
\begin{equation}
	\sigma = \frac{\sqrt{\norm[\big]{\adjoint{\bigl( \mathrm{G}^\prime(\bb{x}_k) \bigr)} \mathrm{G}^\prime(\bb{x}_k)}_{\mathrm{op}}}}{\norm{\FFDop}_{\mathrm{op}}}
\end{equation}
and
\begin{equation}
	\tau = \bigl( \sigma \norm{\FFDop}_{\mathrm{op}}^2 + \norm[\big]{\adjoint{\bigl( \mathrm{G}^\prime(\bb{x}_k) \bigr)} \mathrm{G}^\prime(\bb{x}_k)}_{\mathrm{op}} \bigr)^{-1},
\end{equation}
where \( \norm{}_{\mathrm{op}} \) is the operator norm~\cite[Section 1, (i)]{chambolle2015ergodic}.
These stepsizes satisfy the inequality
\begin{equation}
	\Bigl( \tfrac{1}{\tau} - \norm[\big]{\adjoint{\bigl( G^\prime(\bb{x}_k) \bigr)} G^\prime(\bb{x}_k)}_{\mathrm{op}} \Bigr) \tfrac{1}{\sigma} \geq \norm{\FFDop}_{\mathrm{op}}^2
\end{equation}
needed for convergence, see~\cite[Remark 1]{chambolle2015ergodic}.
We computed \( \norm[\big]{\adjoint{\bigl( G^\prime(\bb{x}_k) \bigr)} G^\prime(\bb{x}_k)}_{\mathrm{op}} \) using \num{10} power iterations and used \( \norm{\FFDop}_{\mathrm{op}} \leq \sqrt{8} \) from~\cite[Theorem 3.1]{chambolle2004algorithm}.
\begin{algorithm}[t]
    \SetAlgoLined%
	\DontPrintSemicolon%
	\KwData{Number of iterations \( K \in \mathbb{N} \), step sizes \( \tau, \sigma > 0 \), initial points \( \bb{w}_0 \in \mathbb{C}^{D} \), \( \bb{z}_0 \in \mathbb{C}^{D \times 2} \)}
	\KwResult{\( \bb{w}_K \), solution to~\eqref{eq:dual subproblem}}
    \For{\( k \leftarrow 0 \) \KwTo \( K - 1 \)}{
		\( \bb{w}_{k + 1} = \bb{w}_k - \tau \Bigl( \adjoint{\FFDop}\bb{z}_k + \adjoint{\bigl( \mathrm{G}^\prime(\bb{x}_k) \bigr)} \bigl( \mathrm{G}^\prime(\bb{x}_k) (\bb{w}_k - \bb{x}_k) + \mathrm{G}(\bb{x}_k) - \bb{Y} \bigr) \Bigr) \)\;
		\( \bb{z}_{k + 1} = \proj_{\{ \norm{}_{2, \infty} < \alpha \}} \bigl( \bb{z}_k + \sigma \FFDop (2 \bb{w}_{k + 1} - \bb{w}_k) \bigr) \)\;
    }
	\caption{\CondatVu{} algorithm for solving~\eqref{eq:dual subproblem}}%
	\label{alg:cv}
\end{algorithm}

For the computational and practical experiments, we run the inner solver (conjugate gradient or \CondatVu) for \( K = \num{100} \) iterations from a warm start provided by the outer Gauss-Newton iterates.
A fixed number of iterations makes this a fast, practical algorithm at the cost of an inexact solution of the inner problems.
The convergence of such inexact \gls{pgn} methods was studied in~\cite{porta2024inexact}.
We use \num{8} iterations of proximal Gauss-Newton for the computational and practical experiments, except for BF-pFPM* and BF-pFPM, where we use \num{4}.
The regularization weight \( \alpha \) was set to \num{0.1} for the computational experiments and to \num{9e4} (\elltwo) and \num{1.5e4} (\gls{tv}) for the practical ones.

\section{Additional Reconstruction Results}%
\begin{figure}
    \centering
    \begin{tikzpicture}
        \foreach \reg/\anno [count=\ireg] in {tv/TV, l2/\elltwo, none/None} {
            \pgfmathsetlengthmacro{\myy}{-\ireg*(\imwidth + \ypad)}
            \node [rotate=90] at (1.5cm, \myy) {\anno};
            \foreach \method/\anno [count=\imethod] in {BF-pFPMprime/BF-pFPM*, BF-pFPM/BF-pFPM, DF-pFPM/DF-pFPM, DF-pFPMprime/DF-pFPM*} {
                \pgfmathsetlengthmacro{\myx}{\imethod*(\imwidth + \xpad)}
                \ifthenelse{\ireg=1}{\node at (\myx, -3.3cm) {\anno};}{}
                \coordinate (onn) at (\myx + 1.25cm, \myy - 0.4cm);
                \coordinate (att) at (\myx - .75cm, \myy - 2.25cm);
                \begin{scope}[spy using outlines={rectangle, magnification=3, width=\imwidth/2, height=\imwidth/2, connect spies}]
                    \node at (\myx, \myy) {\includegraphics[width=\imwidth]{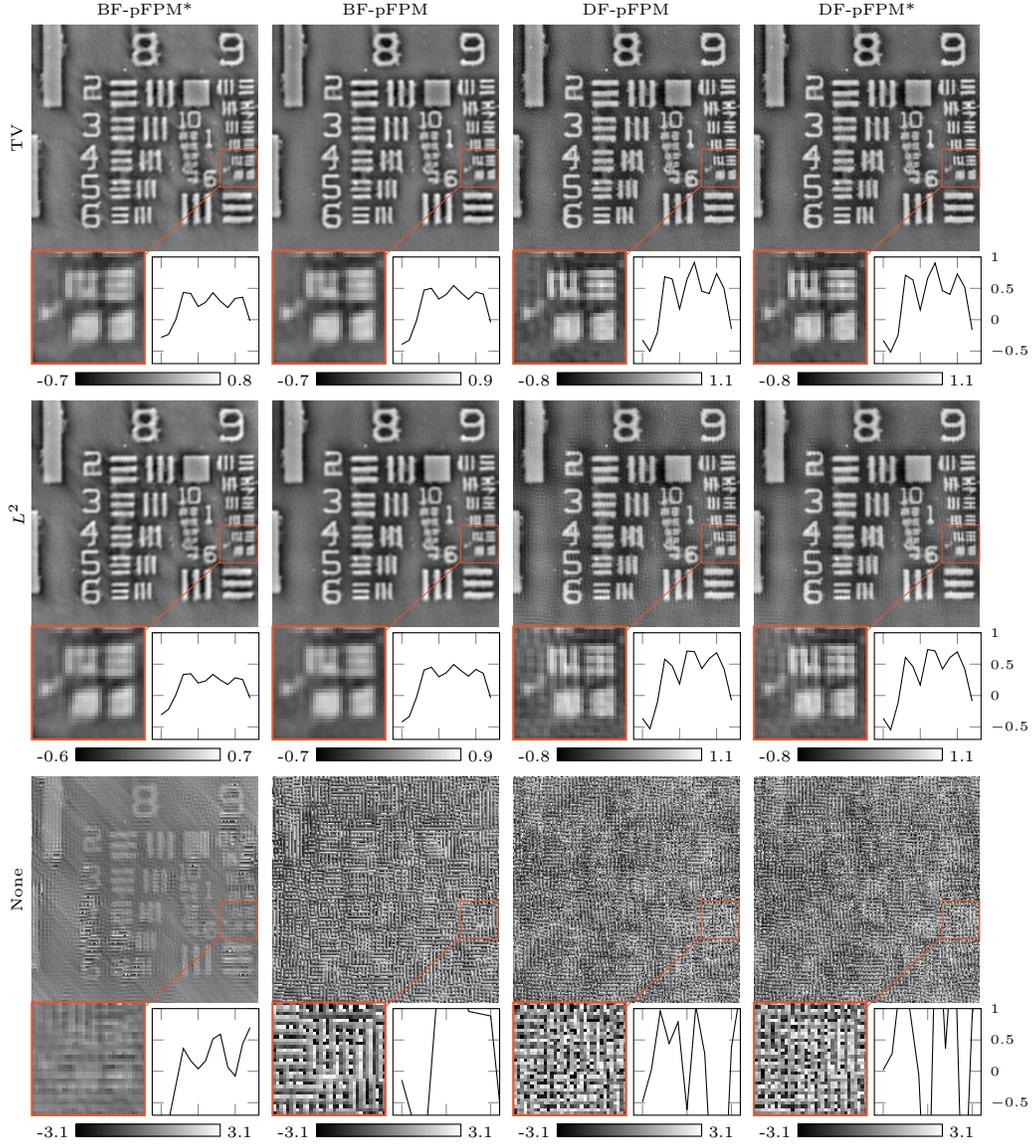}};
                    \spy [spycolor] on (onn) in node at (att);
                \end{scope}
                \node at (\myx, \myy - 3.2cm) {\csvreader[no head]{./results/experiments/\reg/\method/span.csv}{}{\csvcoli\drawcolorbar\ \csvcolii}};
                \begin{axis}[width=3cm, height=3cm, at={(\myx + .1cm, \myy - 3cm)}, yticklabel pos=right, font=\tiny, ymax=1, ymin=-.7, yticklabel={\ifthenelse{\imethod=4}{\( \pgfmathprintnumber{\tick} \)}{\empty}}, xticklabel=\empty]
                    \addplot [mark=none] table [col sep=comma] {./results/experiments/\reg/\method/vals.csv};
                \end{axis}
            }
        }
    \end{tikzpicture}%
    \caption{%
		Influence of the regularizer on the phase reconstruction of the \USAFTarget{}.
        From top to bottom: \gls{tv}; \elltwo{}; no regularization.
        From left to right: BF-pFPM*; BF-pFPM; DF-pFPM; DF-pFPM* methods.
		Inset: fifth and sixth element in the ninth group of the phantom.
        Plot: line profile of the fifth element in the ninth group.
        The axes of the plot are shared between the methods.
    }%
	\label{fig:experimental results all}
\end{figure}

\label{sec:additional reconstruction results}
We show in \cref{fig:experimental results all} additional results of our reconstruction algorithm with \elltwo{} and no regularization.
There, BF-pFPM*, BF-pFPM, and DF-pFPM with \gls{tv} regularization are repeated from~\cref{fig:experimental results} to ease the comparison.
In addition to the illumination patterns shown in~\cref{fig:experimental results}, we show reconstruction results from three annuli comprising \glspl{led} with \gls{na} in the intervals \( [1, 1.3]\numaptobj, [1.3, 1.7]\numaptobj \), and \( [1.7, 2]\numaptobj \) (DF-pFPM*).
The superior noise-suppression properties of \gls{tv} are demonstrated by the lack of high-frequency artifacts that are clearly visible in the \elltwo{} reconstruction.
As striking example, in the reconstruction obtained using \elltwo{} regularization, ringing artifacts due to the vertical stripes in the fifth element in the ninth group significantly impede the identification of the horizontal stripes in the same element.
In the reconstruction obtained using \gls{tv} regularization, however, both the vertical stripes and the horizontal stripes in the same element are easily identified.
The improvement from an increase in the number of annuli is negligible, possibly due to model deviation, unaccounted sources of noise, or ill-suited optimization parameters.
The effect of an increase of the number of annuli is investigated for simulated data in~\cref{fig:pattern comparison five} (two annuli) and~\cref{fig:pattern comparison six} (three annuli) where the results are similar.

\begin{figure}
    \centering
    \begin{tikzpicture}
        \pgfmathsetlengthmacro{\myx}{(\imwidth + \xpad + 7mm)/2+1cm}
        \draw [thick, gray] (\myx, -3.3cm) -- ++(6, 0) node [midway, above, black] {Example measurements};
        \pgfmathsetlengthmacro{\myx}{3*(\imwidth + \xpad + 7mm)}
        \node at (\myx, -3.3cm) {FPM reconstruction};
        \pgfmathsetlengthmacro{\myy}{-1*(\imwidth + \ypad)}
        \foreach \method [count=\imethod] in {brightfield, central_led, FPM} {
            \pgfmathsetlengthmacro{\myx}{\imethod*(\imwidth + \xpad + 7mm)}
            \coordinate (onn) at (\myx + 1.25cm, \myy - 0.4cm);
            \coordinate (att) at (\myx - .75cm, \myy - 2.25cm);
            \begin{scope}[spy using outlines={rectangle, magnification=3, width=\imwidth/2, height=\imwidth/2, connect spies}]
                \node at (\myx, \myy) {\includegraphics[width=\imwidth]{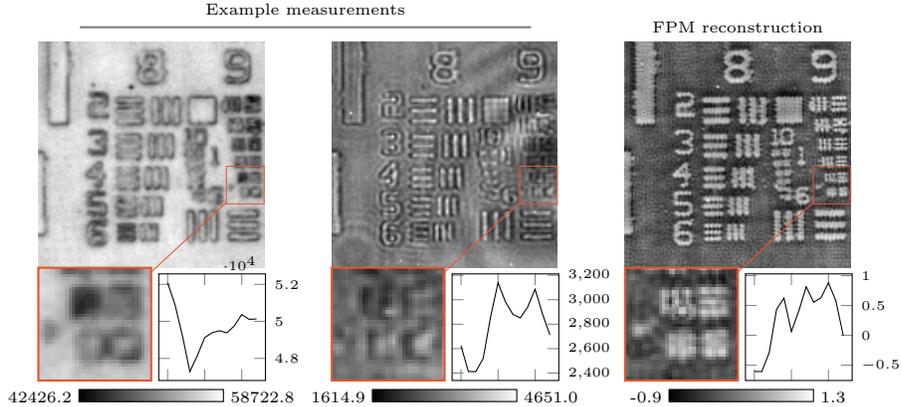}};
                \spy [spycolor] on (onn) in node at (att);
            \end{scope}
            \node at (\myx, \myy - 3.2cm) {\csvreader[no head]{./results/experiments/\method/span.csv}{}{\csvcoli \drawcolorbar\ \csvcolii}};
            \begin{axis}[width=3cm, height=3cm, at={(\myx + .1cm, \myy - 3cm)}, yticklabel pos=right, font=\tiny, xticklabel=\empty]
                \addplot [mark=none] table [col sep=comma] {./results/experiments/\method/vals.csv};
            \end{axis}
        }
    \end{tikzpicture}%
    \caption{%
		\USAFTarget{}.
        From left to right: illumination by all \brightfield{} \glspl{led}; illumination by the central \gls{led}; reconstruction obtained from \gls{fpm}.
        Plot: line profile of the fifth element in group \num{9}.
    }%
	\label{fig:brightfield etc}
\end{figure}

In addition to the reconstruction results from our algorithm, the measurement resulting from all \brightfield{} \glspl{led}, only the central \gls{led}, and the reconstruction from \gls{fpm} data are shown in~\cref{fig:brightfield etc}.
The \gls{fpm} reconstruction was obtained by running
\begin{equation}
	\bb{x}_{k + 1} = \bb{x}_k - \tau \adjoint{A}\bigl( A\bb{x}_k - A\bb{x}_k \oslash (\abs{A\bb{x}_k} + \epsilon) \odot \sqrt{\bb{Y}} \bigr)
\end{equation}
for \num{50} iterations using \( \tau = \num{1e-2} \) and \( \epsilon = \num{1.5e1} \) starting from \( \bb{x}_0 = \mathbf{1} \), where $\oslash$ denotes an element-wise division.
This constitutes gradient descent on the \enquote{amplitude} loss
\begin{equation}
	\bb{x} \mapsto \Half \norm[\big]{\,\abs{A\bb{x}} - \sqrt{\bb{Y}}\,}_2^2,
\end{equation}
which typically is preferred for \gls{fpm} reconstruction \cite{yeh2015experimental}, where \( \epsilon > 0 \) ensures numerical stability.
Here, \( A: \bb{x} \mapsto (\bb{A}_1\bb{x}, \bb{A}_2\bb{x}, \dotsc, \bb{A}_{\Nleds}\bb{x}) \) summarizes the linear operators that correspond to the individual \glspl{led}, and the measurements were acquired with \qty{1}{\second} exposure time.
The \gls{fpm} reconstruction suffers from high-frequency artifacts, again possibly caused by movements during the image acquisition.
For an exposure time of one \qty{1}{\second}, the acquisition of the \gls{fpm} data with \num{126} \glspl{led} takes \qty{126}{\second}.
This outlines the benefit of our approach, which only takes \qty{550}{\milli\second}, thus reducing the chance of movement artifacts which are common in biological applications.

\section{Comparison of Dark-Field Illumination Patterns}%
\label{sec:illumination comparison}
\begin{figure}
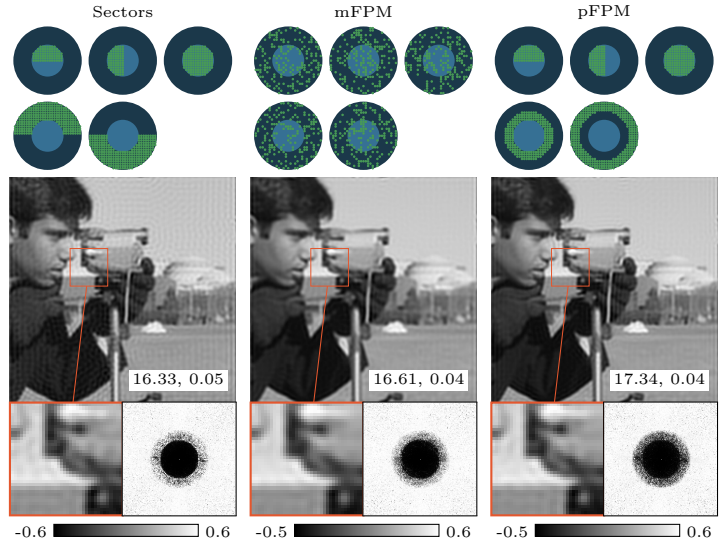

    \centering
	\begin{tikzpicture}
		\pgfmathsetlengthmacro{\wbt}{\imwidth/2}
		\foreach [count=\inpat] \numpat in {5}{
			\pgfmathsetlengthmacro{\myy}{-7.4cm * \inpat}
			\foreach \method/\anno [count=\imethod] in {cone/Sectors, mFPM/mFPM, pFPM/pFPM}
			{
				\pgfmathsetlengthmacro{\myx}{\imethod*(\imwidth + \xpad)}
				\ifthenelse{\inpat=1}{\node at (\myx, -3.7cm) {\anno};}{}
				\coordinate (onn) at (\myx - .45cm, \myy + 0.3cm);
				\coordinate (att) at (\myx - .75cm, \myy - 2.25cm);
				\begin{scope}[spy using outlines={rectangle, magnification=3, width=\imwidth/2, height=\imwidth/2, connect spies}]
					\node at (\myx, \myy) {\includegraphics[width=\imwidth]{./results/simulation/pattern-comparison/\numpat/\method/x_est.png}};
					\spy [spycolor] on (onn) in node at (att);
				\end{scope}
				\pgfmathsetlengthmacro{\wbf}{1cm}
				\foreach [count=\iipat from 0] \ipat in {1,...,\numpat} {
					\pgfmathsetlengthmacro{\myyy}{greater(\ipat, 3) * -(\wbf + 0cm) + \myy + 3.05cm}
					\pgfmathsetlengthmacro{\myyx}{\myx + (mod(\iipat, 3) - 1) * (\wbf + 0cm)}
\node at (\myyx, \myyy) {%
                    \ifthenelse{\ipat=1}{\includegraphics[width=\wbf]{./results/simulation/pattern-comparison/\numpat/\method/patterns/0\iipat.pdf}}{\includegraphics[rotate=180,width=\wbf]{./results/simulation/pattern-comparison/\numpat/\method/patterns/0\iipat.pdf}}
					};
				}
				\node at (\myx, \myy - 3.2cm) {\csvreader[no head]{./results/simulation/pattern-comparison/5/\method/span.csv}{}{\csvcoli\drawcolorbar\ \csvcolii}};
				\node [fill=white, left, inner sep=0.5mm] at (\myx + 1.4cm, \myy - 1.2cm) {\csvreader[no head]{./results/simulation/pattern-comparison/\numpat/\method/metrics.csv}{}{\csvcoli, \csvcolii}};
				\pgfmathsetlengthmacro{\wbt}{\imwidth/2}
				\node at (\myx + .75cm, \myy - 2.25cm) {\includegraphics[width=\wbt,cframe=black]{./results/simulation/pattern-comparison/5/\method/ft_error.png}};
		}
		}
	\end{tikzpicture}
    \caption{%
        Reconstructions from different designs of \( M = 5 \) illumination patterns (left to right: sectors, random, proposed annuli).
		Top: illumination patterns.
		Middle and inset: reconstruction.
		Bottom: relative Fourier error map.
        The numbers in the inset are SNR, NMSE.
        The colormap applies only to the reconstruction.
    }%
    \label{fig:pattern comparison five}
\end{figure}

\begin{figure}
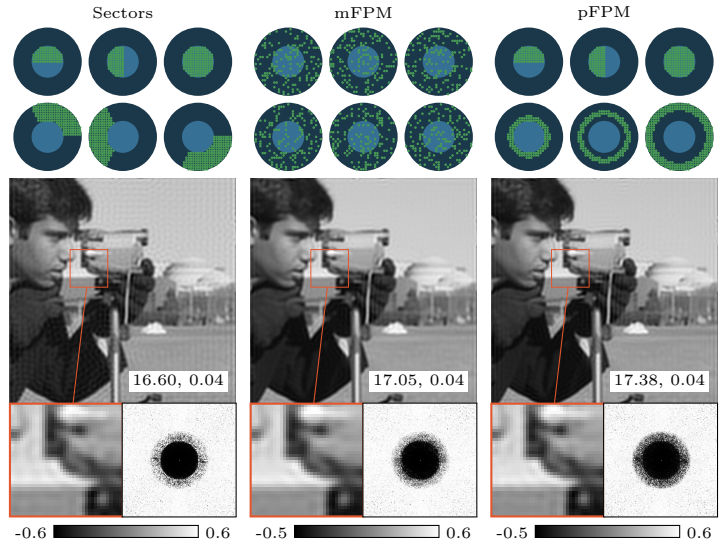

    \centering
	\begin{tikzpicture}
		\pgfmathsetlengthmacro{\wbt}{\imwidth/2}
		\foreach [count=\inpat] \numpat in {6}{
			\pgfmathsetlengthmacro{\myy}{-7.4cm * \inpat}
			\foreach \method/\anno [count=\imethod] in {cone/Sectors, mFPM/mFPM, pFPM/pFPM}
			{
				\pgfmathsetlengthmacro{\myx}{\imethod*(\imwidth + \xpad)}
				\ifthenelse{\inpat=1}{\node at (\myx, -3.7cm) {\anno};}{}
				\coordinate (onn) at (\myx - .45cm, \myy + 0.3cm);
				\coordinate (att) at (\myx - .75cm, \myy - 2.25cm);
				\begin{scope}[spy using outlines={rectangle, magnification=3, width=\imwidth/2, height=\imwidth/2, connect spies}]
					\node at (\myx, \myy) {\includegraphics[width=\imwidth]{./results/simulation/pattern-comparison/\numpat/\method/x_est.png}};
					\spy [spycolor] on (onn) in node at (att);
				\end{scope}
				\pgfmathsetlengthmacro{\wbf}{1cm}
				\foreach [count=\iipat from 0] \ipat in {1,...,\numpat} {
					\pgfmathsetlengthmacro{\myyy}{greater(\ipat, 3) * -(\wbf + 0cm) + \myy + 3.05cm}
					\pgfmathsetlengthmacro{\myyx}{\myx + (mod(\iipat, 3) - 1) * (\wbf + 0cm)}
\node at (\myyx, \myyy) {%
                    \ifthenelse{\ipat=1}{\includegraphics[width=\wbf]{./results/simulation/pattern-comparison/\numpat/\method/patterns/0\iipat.pdf}}{\includegraphics[rotate=180,width=\wbf]{./results/simulation/pattern-comparison/\numpat/\method/patterns/0\iipat.pdf}}
					};
				}
				\node at (\myx, \myy - 3.2cm) {\csvreader[no head]{./results/simulation/pattern-comparison/5/\method/span.csv}{}{\csvcoli\drawcolorbar\ \csvcolii}};
				\node [fill=white, left, inner sep=0.5mm] at (\myx + 1.4cm, \myy - 1.2cm) {\csvreader[no head]{./results/simulation/pattern-comparison/\numpat/\method/metrics.csv}{}{\csvcoli, \csvcolii}};
				\pgfmathsetlengthmacro{\wbt}{\imwidth/2}
				\node at (\myx + .75cm, \myy - 2.25cm) {\includegraphics[width=\wbt,cframe=black]{./results/simulation/pattern-comparison/5/\method/ft_error.png}};
		}
		}
	\end{tikzpicture}
    \caption{%
        Reconstructions from different designs of \( M = 6 \) illumination patterns (left to right: sectors, random, proposed annuli).
		Top: illumination patterns.
		Middle and inset: reconstruction.
		Bottom: relative Fourier error map.
        The numbers in the inset are SNR, NMSE.
        The colormap applies only to the reconstruction.
    }%
    \label{fig:pattern comparison six}
\end{figure}
The proposed annular illumination patterns are inspired by the perturbative principle.
To design them, we compare reconstructions obtained from a variety of common illumination patterns qualitatively and quantitatively, over the same simulation setup as in the main text.
We considered illumination patterns where three \brightfield{} measurements are augmented by \( (\Npatterns - 3) \) annuli with equal annular radius in the \darkfield{}.
Then we fixed the number of illumination patterns to \( M = 5 \) and \( 6 \) and compared our illumination patterns against two designs:

The first design was proposed in~\cite{Xue:19} in the context of data-driven recovery, where the classic two \gls{dpc} half-circles in the \brightfield{} are augmented by annular sectors in the \darkfield{}.
Unfortunately, in the absence of an additional \brightfield{} measurement we observed the same oriented streaking artifacts as in~\cref{fig:simulation result}.
Therefore, we use the same three \brightfield{} illumination patterns as for our method and, consequently, partition the \darkfield{} annulus into \( (\Npatterns - 3) \) equisized sectors.

The second design instantiates a random selection similar to the selection proposed by~\cite{tian2014multiplexed} in the original publication on \gls{mfpm}.
There, in each illumination pattern, a fraction of \( \tfrac{1}{\Npatterns} \) of all \glspl{led} up to \( \numaptill \) is activated.

We conclude from \cref{fig:pattern comparison five} (\( M = 5 \)) and \cref{fig:pattern comparison six} (\( M = 6 \)) that the proposed annular illumination patterns consistently result in superior reconstructions, as seen qualitatively in the Fourier error maps and quantitatively in the \gls{snr} and the \gls{rmse}.
In addition, an increase in the number of measurements from \num{5} to \num{6} gives only marginal gains, thus making the \gls{pfpm} with \num{5} measurements a fast, practical choice for those applications that necessitate high frame rates.
\section{Simulation Data}%
\label{sec:simulation data}
We consider a region \( \Omega \) of \numproduct{220x220} centered around the pixel index (200, 300) of the the \texttt{cameraman} image, \( \cameraman \), and normalize it by
\begin{equation}
	\cameraman_{\mathrm{norm}} = \frac{\cameraman - \min_{\bb{k} \in \Omega} [\cameraman]_{\bb{k}}}{\max_{\bb{k} \in \Omega} [\cameraman]_{\bb{k}} - \min_{\bb{k} \in \Omega} [\cameraman]_{\bb{k}}} - 0.5.
\end{equation}
Then, we generate the simulation target as the phase encoding of the complex image
\begin{equation}
    \bb{x}^\ast = \exp(\ImagUnit \cameraman_{\mathrm{norm}}).
\end{equation}
\end{document}